\documentclass{aa}  

\usepackage{graphicx}
\usepackage{txfonts}
\usepackage{epsfig}
\usepackage{amsmath}
\usepackage{amssymb}
\usepackage{makecell}
\usepackage{natbib}

\begin{document}

   \title{A correlation between chemistry, polarization and dust properties in the Pipe Nebula starless core FeSt\,1-457}
   \titlerunning{Chemistry, polarization \& grain growth in FeSt 1-457}

   \author{Carmen Ju\'arez
          \inst{1,2}, 
          Josep M. Girart
          \inst{1,3}, 
          Pau Frau
          \inst{4},
          Aina Palau
          \inst{5}, 
          Robert Estalella
          \inst{2}, 
          Oscar Morata
          \inst{6}, 
          Felipe O. Alves
          \inst{7}, 
          Maria T. Beltr\'an
          \inst{8}, 
          Marco Padovani
          \inst{8,9}
          }
	\authorrunning{Ju\'arez et al.}
   \institute{$^1$Institut de Ci\`encies de l'Espai, (CSIC-IEEC),
              Campus UAB, Carrer de Can Magrans, S/N, 08193 Cerdanyola del Vall\`es, Catalonia, Spain\\
              \email{juarez@ice.cat}\\
              $^2$Dept. de F\'{\i}sica Qu\`antica i Astrof\'{\i}sica (formerly Astronomia i Meteorologia), Institut de Ci\`encies del Cosmos (ICCUB)\thanks{The ICCUB is a CSIC-Associated Unit through the Institut de Ci\`encies de l'Espai (ICE).}, Universitat de Barcelona (IEEC-UB), Mart\'{\i} Franqu\`es 1, E08028 Barcelona, Spain\\
              $^3$Harvard-Smithsonian Center for Astrophysics, 
              60 Garden St., Cambridge, MA 02138, USA\\
              $^4$Observatorio Astron\'omico Nacional, 
              Alfonso XII 3, 28014 Madrid, Spain\\
              $^5$Instituto de Radioastronom\'ia y Astrof\'isica,
              Universidad Nacional Aut\'onoma de M\'exico, P.O. Box 3-72, 58090, Morelia, Michoacan, Mexico\\
              $^6$Institute of Astronomy and Astrophysics, Academia Sinica,
              P.O. Box 23-141, Taipei 106, Taiwan\\
              $^7$Max-Planck-Institut f$\ddot{u}$r extraterrestrische Physik,
              Giessenbachstrasse 1, D-85748 Garching, Germany\\
              $^8$INAF-Osservatorio Astrofisico di Arcetri,
              Largo E. Fermi 5, 50125 Firenze, Italy\\
             $^9$Laboratoire Univers et Particules de Montpellier, UMR 5299 du CNRS, Universit\'e de Montpellier II,
             place E. Bataillon, cc072, 34095 Montpellier, France
             }


 
  \abstract
   {Pre-stellar cores within molecular clouds provide the very initial conditions in which stars are formed. FeSt\,1-457 is a prototypical starless core and the most chemically evolved among those isolated, embedded in the most pristine part of the Pipe nebula, the bowl.}
   {We use the IRAM 30m telescope and the PdBI to study the chemical and physical properties of the starless core FeSt\,1-457 (Core 109) in the Pipe nebula.}
   {We fit the hyperfine structure of the N$_2$H$^+$ (1$-$0) IRAM 30m data. This allow us to measure with high precision the velocity field, line widths and opacity and derive the excitation temperature and column density in the core. We use a modified Bonnor-Ebert sphere model adding a temperature gradient towards the center to fit the 1.2 mm continuum emission and visual extinction maps. Using this model, we estimate the abundances of the N$_2$H$^+$ and the rest of molecular lines detected in the 30 GHz wide line survey performed at 3 mm with IRAM 30m using ARTIST software.}  
   {The core presents a rich chemistry with emission from early (C$_3$H$_2$, HCN, CS) and late-time molecules (e.g., N$_2$H$^+$), with a clear chemical spatial differentiation for nitrogen (centrally peaked), oxygen (peaking to the southwest) and sulphurated molecules (peaking to the east). For most of the molecules detected (HCN, HCO$^+$, CH$_3$OH, CS, SO, $^{13}$CO and C$^{18}$O), abundances are best fitted with three values, presenting a clear decrease of abundance of at least 1 or 2 orders of magnitude towards the center of the core. The Bonnor-Ebert analysis indicates the core is gravitationally unstable and the magnetic field is not strong enough to avoid the collapse.}
   {Depletion of molecules onto the dust grains occurs at the interior of the core, where dust grain growth and dust depolarization also occurs. This suggests that these properties may be related. On the other hand, some molecules exhibit asymmetries in their integrated emission maps, which appear to be correlated with a previously reported submillimetre polarization asymmetry. These asymmetries could be due to a stronger interstellar radiation field in the western side of the core.}

   \keywords{Stars: formation -- 
   		ISM: individual objects: FeSt 1-457, core 109, Pipe nebula --
                	ISM: molecules --
          	Radiative transfer
               }

   \maketitle
%

\section{Introduction}
   Pre-stellar cores within molecular clouds constitute the stage previous to the star formation process. They provide the very initial conditions in which stars are formed. They are typically cold ($\le10$ K) with temperature gradients decreasing towards the center \citep[e.g.,][]{Ruoskanen11,Wilcock12,Launhardt13}, and densities $>10^4$ cm$^{-3}$, with a chemistry affected by molecular freeze-out onto dust grains \citep[e.g.,][]{Tafalla02,Marsh14}. Their physical structure can be well explained through a Bonnor-Ebert profile  \citep{Bonnor56,Ebert55}, i.e., the profile corresponding to an isothermal gas sphere in hydrostatic equilibrium \citep[e.g.,][]{Evans01,Kandori05,Roy14}. However, their density and temperature gradient structures are still under debate \citep[e.g.,][]{Sipila11,Hardegree-Ullman13} and the implications of heavy molecular depletion on other physical properties, such as dust grain properties and polarization, are not clear.

FeSt\,1-457 \citep{FeSt84}, also known as Core 109 in \citet{Lombardi06} catalogue, is a prototypical starless core, as previous studies have shown that this core is quiescent with no IRAS nor Spitzer Space Telescope point sources associated \citep{Forbrich09,Forbrich15,Ascenso13}. Indeed, the internal luminosity, estimated using the non-detection from the 70 $\mu$m Herschel\footnote{Herschel is an ESA space observatory with science instruments provided by European-led Principal Investigator consortia and with important participation from NASA.} image  ($3\times\mathrm{rms}=0.02$ Jy pixel$^{-1}$) and using the relation between the 70 $\mu$m flux and the internal luminosity \citep{Dunham08}, is L$_\mathrm{int}<0.004$ L$_\odot$. It is also the most chemically evolved starless core among those isolated, embedded in the most pristine part of the Pipe nebula, the bowl, at a distance of 145 pc \citep{AlvesFranco07}. \citet{Frau15} suggest the presence of two filaments with north-south and east-west directions colliding along the northwest-southeast direction at the bowl region, just where FeSt\,1-457 is located.  

\citet{Frau12a} carried out a 15 GHz bandwidth molecular survey at 3 mm towards FeSt\,1-457 with the Institute of Millimeter Radioastronomy (IRAM) 30m telescope. They find a rich chemistry (specially compared with most of the other Pipe starless cores): many lines are detected from early- (C$_3$H$_2$, HCN, CS, C$^{34}$S and CN) and late-time (N$_2$H$^+$, N$_2$D$^+$ and DCO$^+$) molecules, suggesting that FeSt\,1-457 is an evolved starless core. \citet{Aguti07} observed the 3 mm lines of N$_2$H$^+$, HCO$^+$, CS and C$^{18}$O with the IRAM 30m. They find evidence of CO and HCO$^+$ depletion in the center of the core. However, the N$_2$H$^+$ ($1-0$) may be undepleted to depths of $\la40$ mag of visual extinction.

FeSt\,1-457 has a mass of $\sim4$ M$_\odot$ \citep{Frau10,Roman-Zuniga10}, quasi spherical and compact structure and shows signs of gravitational instability \citep{Kandori05,Frau10}. \citet{Aguti07} find that FeSt\,1-457 might be pulsating, based on expansions of the outer layers. However, their Jeans mass measurement is compatible with the mass of the core and they propose a quasi-stable state near hydrodynamic equilibrium. 

The core is embedded in a magnetised medium \citep{Alves08,Alves14,Franco10}, which suggests that the magnetic field could be a source of external support. \citet{Alves14}, in a multi scale polarization study towards FeSt\,1-457, find a polarization `hole' for radii $\la55$ arcsec which they propose to be the result of loss of grain alignment with the magnetic field due to the lack of an internal source of radiation. 


In this work we study the starless core FeSt\,1-457 through N$_2$H$^+$ (1$-$0) observations obtained with the IRAM 30m and Plateau de Bure Interferometer (PdBI) telescopes\footnote{IRAM is supported by INSU/CNRS (France), MPG (Germany) and IGN (Spain).}. N$_2$H$^+$ is known to be a very good tracer of the densest inner parts of dense cores as it is not expected to be depleted until densities >$10^5$ cm$^{-3}$ \citep[e.g.,][]{Bergin07}. In addition, we present mapping information of 15 molecular lines detected at 3 mm within a frequency range spanning 30 GHz obtained with the IRAM 30m which allowed us also to perform a detailed study of the abundance radial profiles and to relate them to the known grain growth and depolarization in the center.

\section{Observations}
\subsection{IRAM 30m} 
\subsubsection{Line observations}
%
   %
\begin{table*}
\caption{FTS molecular transitions}
\centering
\begin{tabular}{c c c c c c}
\hline
\hline
Molecule&Transition&Frequency &HPBW$^\mathrm{a}$&E$_{\mathrm{U}}^\mathrm{b}$&$\rho_{\mathrm{crit}}^\mathrm{c}$\\
  	      &		       &(GHz)         &(arcsec)                       &(K) 						&(cm$^{-3}$)\\ 
\hline
 U$^\mathrm{d}$	&$-$						&80.5599		&30.5	&$-$		&$-$	\\
 HC$_3$N			&(9$-$8)					&81.8816		&30.0	&19.6	&$4.2\times10^5$	\\
 C$_3$H$_2$		&($2_{0,2}$$-$$1_{1,1}$)	 	&82.0935		&30.0	&6.4		&$-$	\\
 C$_3$H$_2$		&($2_{1,2}$$-$$1_{0,1}$) 		&85.3388		&28.8	&6.4		&$-$	\\
 C$_4$H			&($9-$8, $J$=17/2$-$15/2)	&85.6725		&28.7	&4.1		&$-$	\\
 NH$_2$D			&($1_{1,1}$$-$$1_{0,1}$) 		&85.9262		&28.6	&20.7	&$-$	\\
 H$^{13}$CN		&(1$-$0)					&86.3399		&28.5	&4.1		&$2.0\times10^6$	\\
 HCN			&(1$-$0)					&88.6318		&27.8	&4.2		&$2.2\times10^6$	\\
 HCO$^+$			&(1$-$0)					&89.1885		&27.6	&4.3		&$1.6\times10^5$	\\
 HNC			&(1$-$0)					&90.6635		&27.1	&4.4		&$2.8\times10^5$	\\
 N$_2$H$^+$		&(1$-$0)					&93.1734		&26.4	&4.5		&$1.4\times10^5$	\\
 C$^{34}$S		&(2$-$1)					&96.4129		&25.5	&6.2		&$-$	\\
 CH$_3$OH		&($2_{-1,2}$$-$$1_{-1,1}$)	&96.7393		&25.4	&12.5	&$-$	\\
 CH$_3$OH		&($2_{0,2}$$-$$1_{0,1}$)		&96.7413		&25.4	&7.0		&$-$	\\
 $^{34}$SO		&($3_2$$-$$2_1$)			&97.7153		&25.2	&9.1		&$-$\\
 CS				&(2$-$1)					&97.9809		&25.1	&7.1		&$3.3\times10^5$	\\
 SO				&($3_2$$-$$2_1$)			&99.2998		&24.8	&9.2		&$-$	\\
 HC$_3$N			&(11$-$10)				&100.076		&24.6	&28.8	&$8.8\times10^5$	\\
 C$^{18}$O		&(1$-$0)					&109.7821	&22.3	&5.3		&$1.9\times10^3$	\\
 U$^\mathrm{d}$	&$-$						&109.9531	&22.4	&$-$		&$-$	\\
 $^{13}$CO		&(1$-$0)					&110.2013	&22.3	&5.3		&$1.9\times10^3$	\\
 \hline
\end{tabular}
\label{tab:transitions}
\begin{list}{}{}
\item[$^\mathrm{a}$]$[\text{HPBW/arcsec}]=2460\times[\text{freq/GHz}]^{-1}$.
\item[$^\mathrm{b}$]E$_{\mathrm{U}}$ is the energy of the upper level of the transition.
\item[$^\mathrm{c}$] $\rho_{\text{crit}}$=A$_{\text{ul}}/\gamma$, where A$_{\text{ul}}$ is the Einstein spontaneous emission coefficient and $\gamma$=$\sigma<$v$>$ is the collisional rate where $\sigma$ is the cross section of the collision for each transition and $<$v$>\approx$(3kT/m)$^{1/2}$ is the average velocity of the collisional particles. As H$_2$ is the most abundant molecule it is used as the dominant collisional particle. We assume T=10 K. The Einstein spontaneous emission coefficients and $\gamma$ values are taken from LAMDA database (http://home.strw.leidenuniv.nl/$\sim$moldata/).
\item[$^\mathrm{d}$]Unidentified line.
\end{list}
\end{table*}
The observations were carried out on July 2012 at the IRAM 30m telescope in Granada, Spain. We used the Eight MIxer Receiver (EMIR) in the E090 configuration.  The backend used was the FTS spectrometer with a channel resolution of 195 kHz (0.6 km~s$^{-1}$ at 3 mm), which provided a total bandwidth of 30 GHz covering the frequency ranges from 78.6 to 86.4 GHz, from 87.6 to 102.0 GHz and from 103.2 to 111.0 GHz (see Table~\ref{tab:transitions}).  In addition, the VESPA (Versatile Spectrometer Array) correlator was used simultaneously with the FTS in order to observe the N$_2$H$^+$ (1$-$0) line with a better channel resolution, 20 kHz (corresponding to 0.063~km~s$^{-1}$), with a total bandwidth of 40~MHz.  The observations were done using the frequency-switching mode with a frequency throw of $\sim$7~MHz. The maps were done in on-the-fly (OTF) mode covering a region of $4.5$~arcmin~$\times$~4.5~arcmin centered at RA(J2000)$=17^{\rm h} 35^{\rm m} 47\fs70$, Dec(J2000)$=-25^{\circ}32'52\farcs9$. In the OTF mode, the maps were done with scans either along the right ascension or either along the declination (with Nyquist separation between scans). This allows to have a better sampling on the observed region. The maps were combined and regrided to have a final pixel size of 6~arcsec.  Pointing and focusing were done following the recommended procedure at the telescope, which should give a pointing accuracy of $\sim$2~arcsec.  Absolute calibration accuracy with EMIR is better than 10\%. In this paper we present the spectra in main beam brightness temperature scale, which was obtained by using the efficiencies reported on the IRAM website. The weather conditions were good with system temperatures of 125--192 K and zenith opacities around 0.07.  The reduction of the final maps and most of the figures have been done with the {\sc{gildas}}\footnote{http://www.iram.fr/IRAMFR/GILDAS/} software. Given the highly distorted baselines generated in the frequency switching mode, the spectral baselines were subtracted only over a small frequency range around a given line. The typical rms noise per channel for the VESPA spectral map is $\Delta T_{\rm mb} \simeq 0.18$~ K. The half power beam width (HPBW) for each frequency is given in Table~\ref{tab:transitions}. 

\subsubsection{Continuum observations}
FeSt\,1-457 was observed on April and May 2009 and on January 2010 with the 117-receiver Max-Planck Millimetre Bolometer (MAMBO-II) of the IRAM 30m telescope (previously described in \citet{Frau10}). 

\subsection{PdBI}
The PdBI observations of N$_2$H$^+$ (1$-$0) were performed on several runs in October and November 2011, and March 2012, in the C array configuration, which provides a baseline range from 24~m to 176~m (7.5--55 k$\lambda$ at 3.2~mm). The observations were done in mosaic mode with an hexagonal pattern of 12 pointings and with a separation of $\sim$18~arcsec. This allows to uniformly cover an area of  $\sim102$$\times102$~arcsec$^2$ centered on RA(J2000)$=17^h35^m47\fs804$, Dec(J2000)$=-25^{\circ}33'01\farcs10$. The primary beam at 3.2 mm is $\sim54$ arcsec. The quasars J1733$-$130 and J1743$-$169 were used as gain calibrators and 3C273, J2015+371, 3C454.3 and J1751+096 as bandpass calibrators. Absolute flux calibration was performed on MWC349 and on the suitable solar system object available at the time of the observations.  The absolute flux accuracy is $\sim$10\%.  The N$_2$H$^+$ (1$-$0) line was observed in a spectral window of 20~MHz with 512 channels, which provides a spectral resolution of 0.039~MHz (0.125~km~s$^{-1}$). The system temperatures and the precipitable water vapor values were between $\sim100$--300 K and $\sim2$--6 mm respectively.   The data were reduced using {\sc{clic}} and {\sc{mapping}} packages of the {\sc{gildas}} software using the standard procedures. The rms noise per channel achieved was 25 mJy~beam$^{-1}$. The synthesized beam was 10.6$\times$ 5.7~arcsec$^2$, with a P.A. of 11.94$^{\circ}$.

\subsection{PdBI+IRAM 30m} \label{combined_obs}
We combined the N$_2$H$^+$ single dish data taken with the IRAM 30m telescope (with an rms noise per channel of $\sim180$ mK) with the PdBI data (rms noise per channel of $\sim53$ mK) to add the short-spacing information not detected by the millimetre interferometer. We used the short-spacing processing widget of the {\sc{mapping}} package of the {\sc{gildas}} software which produces visibilities from the single dish observations with the PdBI spectral resolution and merges them with the PdBI data. After trying different single dish weight factors, we found the best value to be 0.1. This value was the best compromise between adding excessive noise from the PdBI data due to the weak signal and giving to much weight to the single dish data not being able to see the contribution of the interferometer data.

When creating the dirty map we used a $uv$-taper of 65~m (20.3 k$\lambda$) to avoid the noise contribution of the longer $uv$-distances, recovering the highest signal-to-noise ratio. For the cleaning process we used the Steer, Dewdney and Ito (SDI) algorithm \citep{Steer84} to improve some strips that appeared using the H\"ogbom algorithm due to the extended emission. The velocity resolution of the channel maps was 0.12 km s$^{-1}$ as in the PdBI observations and the cutoff value to stop the cleaning was 1.5 times the rms, 0.06 Jy beam$^{-1}$. The synthesized beam was $10.78$~arcsec~$\times$~6.11~arcsec, with a P.A. of 11$^{\circ}$.

\subsection{Extinction data}
We have made use of the high angular resolution extinction maps by \citet{Roman-Zuniga09,Roman-Zuniga10}. These maps were constructed from a concerted deep near-infrared imaging survey using several telescopes (ESO-VLT, ESO-NTT and CAHA 3.5 m) as well as the 2MASS data. These maps have a resolution three times higher than the previous extinction map of this cloud by \citet{Lombardi06}, allowing to resolve structures down to $\sim2000$ AU.

\section{Results}
\subsection{N$_2$H$^+$ (1$-$0)} 
\subsubsection{VESPA IRAM 30m} \label{N2Hp30m}

\begin{figure*}
 \centering
 \includegraphics[scale=0.6,keepaspectratio=true]{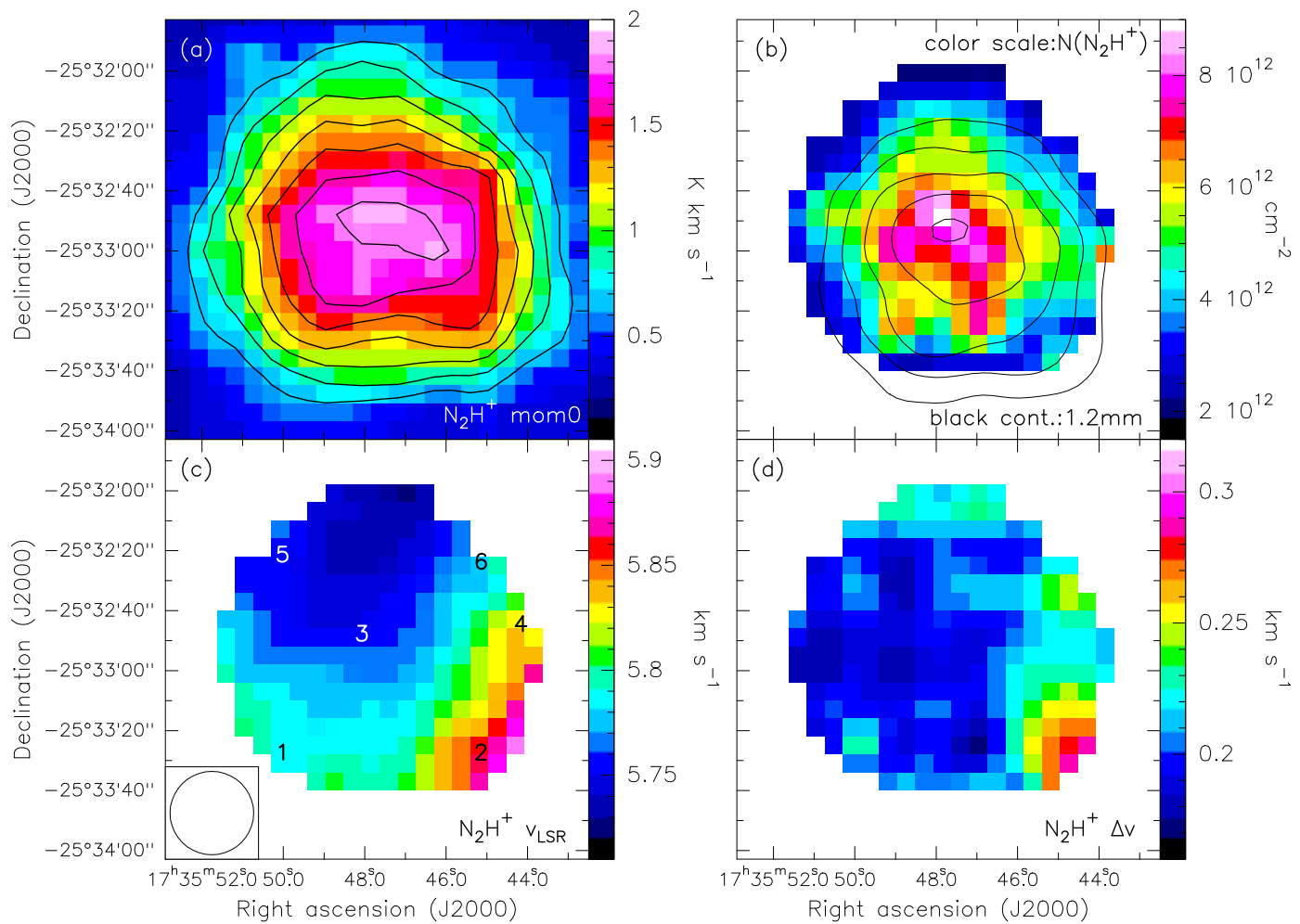}
  \caption{N$_2$H$^+$ (1$-$0) IRAM 30m VESPA data: (a) Moment-zero map (using the three central hyperfine lines) of the IRAM 30m data. Contours are 35, 45, ..., 95 percent of the value of the peak, 1.9 K km s$^{-1}$. (b) Color scale: column density map. Black contours: IRAM 30m MAMBO-II 1.2 mm dust continuum emission at an angular resolution of 21~arcsec.  Contour levels are 30, 40, 50, 60, 70 and 80 percent of the peak value, 122 mJy~beam$^{-1}$. (c) Velocity map and (d) line width map resulting from the hyperfine fit. Numbers in panel (c) indicate the position of the fitted spectra shown in Fig.~\ref{fig:hfsfits} and Table~\ref{tab:hfsfitex}. The HPBW of 27.8 arcsec is shown at bottom left corner of panel (c).}
  \label{fig:VESPAhfs}
\end{figure*}

In Figure~\ref{fig:VESPAhfs}a we present the moment-zero map (i.e. velocity integrated intensity map) of the N$_2$H$^+$ line emission observed with the IRAM 30m telescope. The moment-zero emission was computed in the $4-8$ km s$^{-1}$ velocity range, which covers the three central hyperfine lines. The structure is compact with a size of $\sim~2$ arcmin ($\sim19000$ au) and the intensity increases towards the center of the map up to 2 K km s$^{-1}$. 

In order to infer the physical and chemical properties in FeSt\,1-457 from the N$_2$H$^+$ (1$-$0) IRAM 30m data, we used the HfS tool\footnote{http://ascl.net/1607.011} \citep{Estalella16} to fit the hyperfine structure of our spectra. With the HfS\_fit\_cube procedure we could fit all the spectra of our data cube pixel-by-pixel. The fit was done only for the spectra with a peak intensity larger than five times the rms noise. We obtained the physical parameters $v_\mathrm{LSR}$, $\Delta v$ and $\tau$ and derived the excitation temperature $T_\mathrm{ex}$ and the column density $N$ following \citet{Caselli02b}. 

\begin{figure*}
 \centering
 \includegraphics[scale=0.6,keepaspectratio=true,trim=3.2cm 0 0 2.2cm,clip]{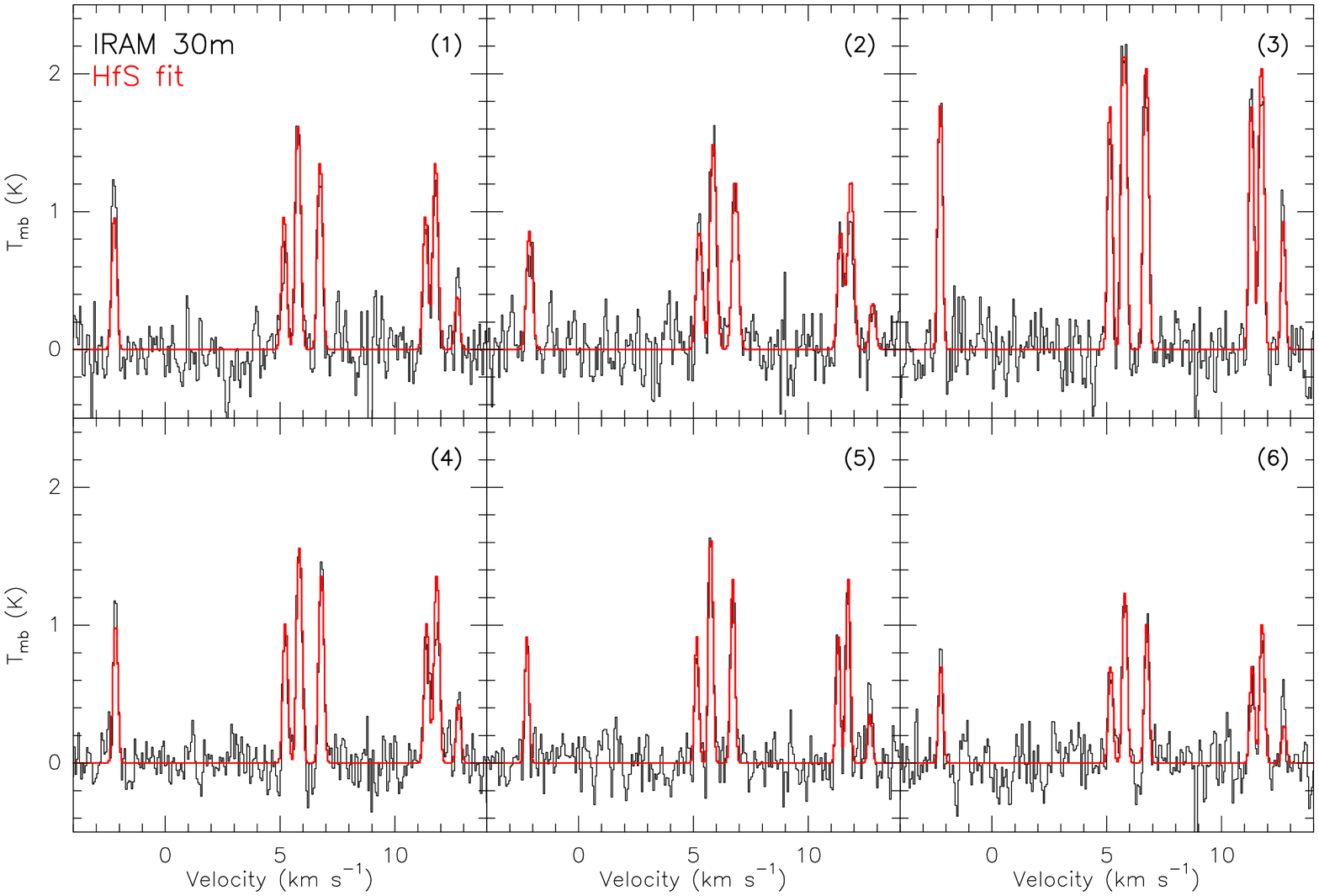}
  \includegraphics[scale=0.6,keepaspectratio=true,trim=3.2cm 0 0 2.0cm,clip]{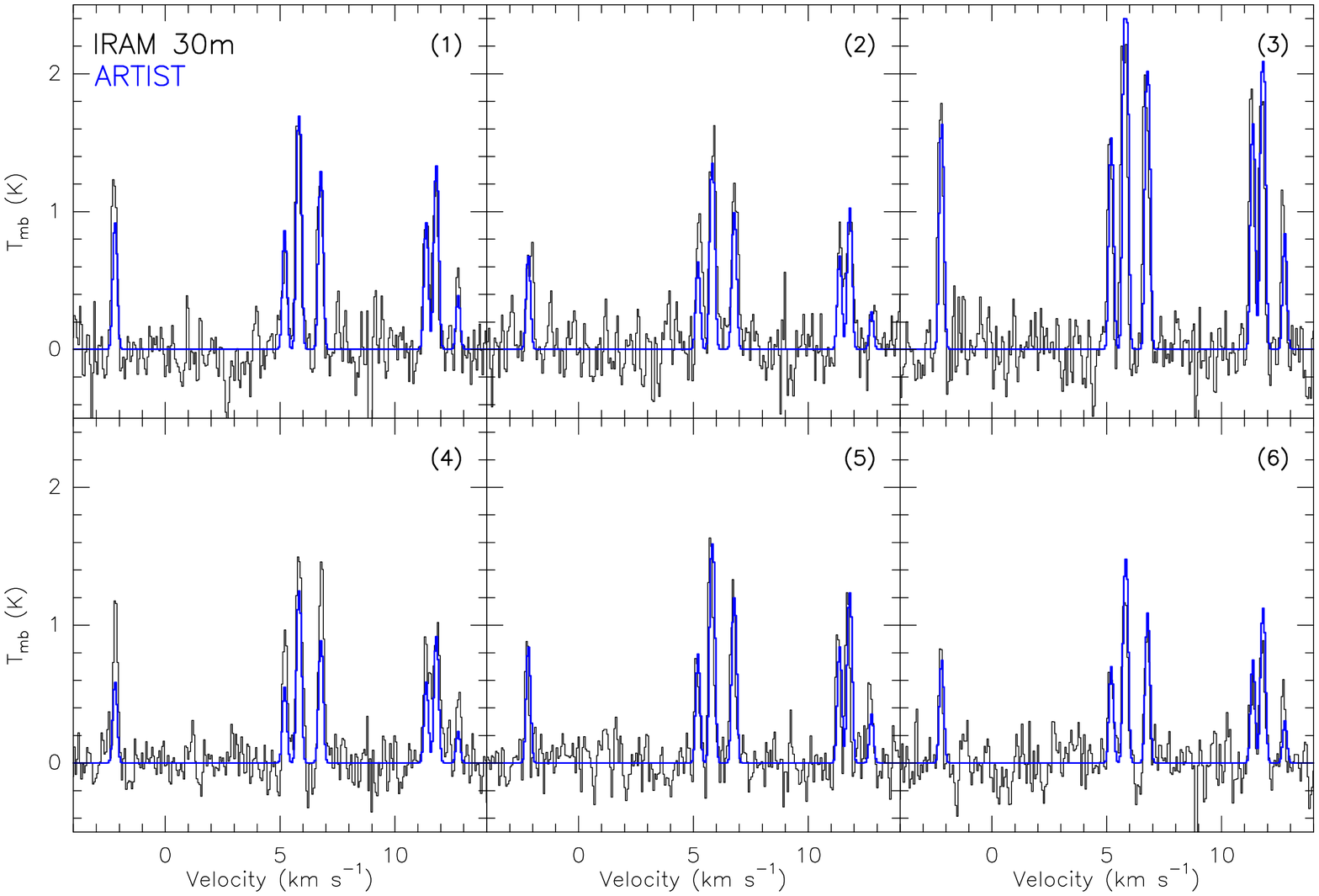}
  \caption{{\bf Upper pannels:} HfS fit of the hyperfine structure of N$_2$H$^+$ (1$-$0) IRAM 30m spectral data at different positions of the core (labeled in Fig.~\ref{fig:VESPAhfs}c). Black and red lines are the observational data and the fit, respectively. {\bf Lower panels:} ARTIST simulated N$_2$H$^+$ (1$-$0) spectral data (see Section~\ref{artist}). Black and blue lines are the observational and simulated data, respectively. Note that the simulated core from ARTIST is spherically symmetric. The discrepancies between the observational and the simulated spectra in panel (4) may be due to the slightly elongated shape of the core towards the west.} 
  \label{fig:hfsfits}
\end{figure*}
\begin{table*}
\caption{HfS fit parameters at different positions of FeSt\,1-457.}
\begin{center}
{\small
\begin{tabular}{cccccccccc}
\noalign{\smallskip}
\noalign{\smallskip}
\hline
\hline
A$\times\tau_m$$^\mathrm{a}$
&$v_\mathrm{LSR}$$^\mathrm{b}$
&$\Delta v$ $^\mathrm{c}$
&
&rms
&Position 
\\
(K)
&(km s$^{-1}$)
&(km s$^{-1}$) 
&$\tau_m$$^\mathrm{d}$
&(K)
&(see Fig.~\ref{fig:VESPAhfs}c)\\
\noalign{\smallskip}
\hline\noalign{\smallskip}
3.1$\pm$0.6 	&5.785$\pm$0.012	&0.231$\pm$0.019	&1.4$\pm$0.4	&0.176   &1\\
2.5$\pm$0.5	&5.863$\pm$0.014	&0.277$\pm$0.022	&1.1$\pm$0.4	&0.166   &2\\
8.7$\pm$1.1	&5.752$\pm$0.007	&0.200$\pm$0.009	&4.0$\pm$0.5	&0.166   &3\\	
3.5$\pm$0.4	&5.839$\pm$0.008   &0.220$\pm$0.013   &1.9$\pm$0.3   &0.124   &4\\
2.7$\pm$0.4	&5.755$\pm$0.008	&0.199$\pm$0.014	&1.1$\pm$0.3	&0.130   &5\\
2.2$\pm$0.5	&5.788$\pm$0.014	&0.210$\pm$0.022	&1.2$\pm$0.5	&0.156   &6\\
\hline
\end{tabular}
\begin{list}{}{}
\item[$^\mathrm{a}$]A$\times\tau_m$, product of the amplitude (A$=f[J_\nu(T_\mathrm{ex})-J_\nu(T_\mathrm{bg})]$, where $f$ is the filling factor, $J_\nu(T)$ is the Planck function in units of temperature, $T_\mathrm{ex}$ is the excitation temperature and $T_\mathrm{bg}$ is the cosmic microwave background temperature) times the optical depth of the main component.
\item[$^\mathrm{b}$]$v_\mathrm{LSR}$, central velocity of the main component.
\item[$^\mathrm{c}$]$\Delta v$, linewidth of the hyperfine components.
\item[$^\mathrm{d}$]$\tau_m$, optical depth of the main component. \citep[See][for further details of the HfS fitting procedure.]{Estalella16} 
\end{list}
}
\end{center}
\label{tab:hfsfitex}
\end{table*} 
%
%

The procedure used for fitting the hyperfine structure of N$_2$H$^+$ (1--0)
assumes local thermodynamic equilibrium (LTE) for all the seven hyperfine lines
of the transition. However, it has been found that the hyperfine lines F$_1$, F=1,0$\to$1,1 and 1,2$\to$1,2 can be out of LTE. This is the case of the
quiescent core L1512, where these excitation anomalies are attributed to
collisional pumping, if the collisional coefficients of the hyperfines were
different, or to  radiative trapping effects depending on the different optical
depth of the hyperfines \citep{Caselli95,Daniel06}.
Thus, in order to avoid this problem, we performed the fit of the hyperfine
structure of N$_2$H$^+$ (1--0) using only the five hyperfine components without
excitation anomalies. In Fig.~\ref{fig:hfsfits} and Table~\ref{tab:hfsfitex} we show the results from the fit
for six selected positions (shown in Fig.~\ref{fig:VESPAhfs}c) of FeSt 1-457.
As can be seen in Fig.~\ref{fig:hfsfits}, although there are some slight discrepancies in the
amplitudes of the two hyperfine lines out of LTE (the two with the highest
velocity), the synthetic spectra (assuming LTE) fit very well the observed
spectra.

In Figure~\ref{fig:VESPAhfs}b we show the resulting N$_2$H$^+$ column density map overlapped with the 1.2 mm dust continuum emission. The values range from 3$\times$10$^{12}$ cm$^{-2}$ at the edges to 9.0$\times$10$^{12}$ cm$^{-2}$ towards the center, with hints of an arc-like structure near the continuum peak marginally resolved (see Section~\ref{combined} below).

The velocity map shown in Fig.~\ref{fig:VESPAhfs}c presents a very smooth gradient in the northeast-southwest direction. The velocity ranges from 5.72$\pm0.01$~km~s$^{-1}$ at the northeast to 5.90$\pm0.01$~km~s$^{-1}$ at the southwest. Excluding the southwesternmost region, the velocity gradient along the northeast-southwest direction is $\simeq$1.8~km~s$^{-1}$~pc$^{-1}$. \citet{Aguti07} find a similar velocity pattern in the C$^{18}$O (1$-$0) and (2$-$1) channel maps.

The typical line widths (see Fig.~\ref{fig:VESPAhfs}d) are 0.17--0.23 ($\pm0.03$~km~s$^{-1}$), except in the southwesternmost region, where it raises up to 0.30$\pm0.03$ km s$^{-1}$ \citep[also in agreement with the results from][]{Aguti07}. Note that the eastern side of the core shows the lowest line widths. The thermal line width for N$_2$H$^+$ has a value of 0.12 km s$^{-1}$, for a kinetic temperature of 10 K. Thus, the three-dimensional non thermal dispersion ($\sigma_{\mathrm{3D,nth}}$) ranges between 0.09 and 0.21 km s$^{-1}$. The isothermal sound speed value $c_s$ is 0.19 km s$^{-1}$, which gives a Mach number (defined as $\sigma_{\mathrm{3D,nth}}$/$c_s$) between 0.46 to 1.1 \citep[following][]{Palau15}. These values indicate that the non thermal motions seem to be subsonic \citep[as reported by][]{Frau10,Frau12b,Alves14} and the thermal pressure seems to be dominant over the turbulent pressure in the core \citep[in agreement with][]{Aguti07} except at the southwestern corner where the ratio seems to be more balanced.
%
\subsubsection{PdBI data}
\begin{figure*}
 \centering
 \includegraphics[scale=0.6,keepaspectratio=true]{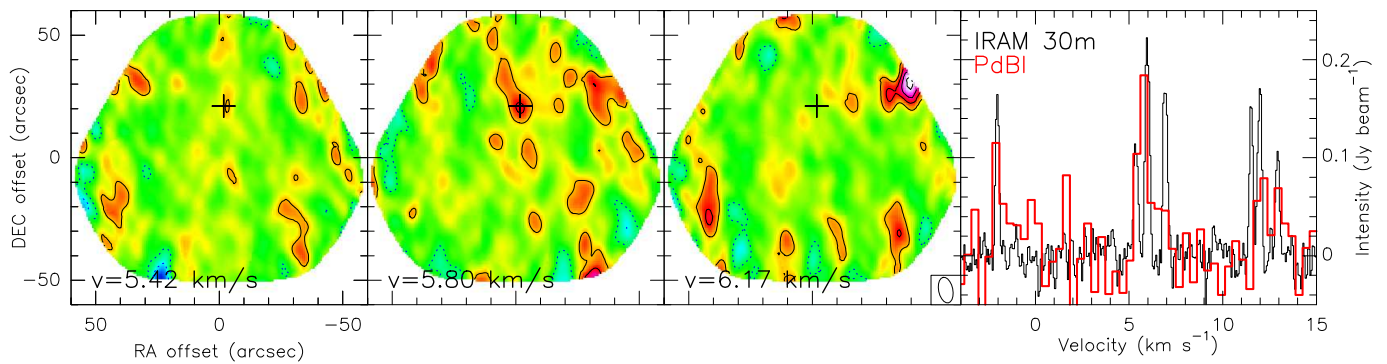}
  \caption{PdBI channel map of N$_2$H$^+$ (1$-$0). Contours are $-$6, $-$3, 3, 6, 9, 12 times the rms level of the map, 25 mJy beam$^{-1}$ ($\sim53$ mK). The synthesized beam located at the bottom right corner of the third panel is $10.6\times5.74$ arcsec$^2$, with a P.A. of 11.94$^{\circ}$. Note that the rms noise increases at the edge of the images. Last panel shows the PdBI spectrum (red line) at the position of the cross. The IRAM 30m N$_2$H$^+$ spectrum (black) has been divided by a factor of 5 to compare the hyperfine structure with the PdBI spectrum.}
  \label{fig:PdB-chmap}
\end{figure*}

In Figure~\ref{fig:PdB-chmap} we show a section of the channel map showing the N$_2$H$^+$ (1$-$0) emission detected with the PdB interferometer. The weak emission ($6\sigma$) is better detected in the second panel corresponding to the strongest line of the three central hyperfine lines, at a velocity of 5.80 km s$^{-1}$. The last panel of Fig.~\ref{fig:PdB-chmap} shows the spectra (in red) at the position of the $6\sigma$ detection. The IRAM 30m spectra (in black) is also shown as comparison, and we can see that the hyperfine structure is also seen in the PdBI spectrum indicating that indeed there is a real N$_2$H$^+$ detection from the PdBI. The emission filtered out by the PdBI with respect to the IRAM 30m is estimated to be $\sim98\%$, from a comparison of the PdBI spectra (convolved with a Gaussian to obtain the IRAM 30m beam size) and the IRAM 30m spectra at the peak position. Given the smallest antenna baseline of PdBI observations, the largest angular scale detectable by the PdBI observations is $\sim12$ arcsec or $\sim1700$ au \citep[following Appendix of][]{Palau10}. Thus, about $98\%$ of the emission detected with the IRAM 30m comes from structures at scales $\ge1700$ au.
\subsubsection{IRAM 30m+PdBI} \label{combined}
\begin{figure}
 \centering
 \includegraphics[scale=0.4,keepaspectratio=true]{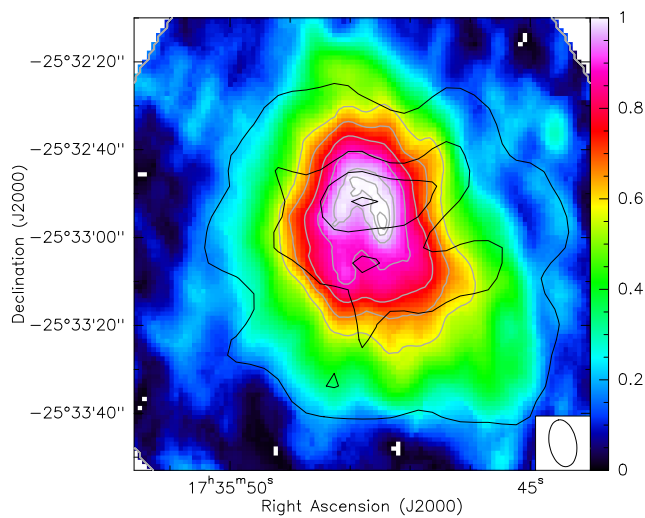}
  \caption{Color scale: Moment-zero map of the IRAM 30m and PdBI combined data. Contours are 57, 67, 77, 88, 97, 99.5, 100 percent of the peak, 1 Jy beam$^{-1}$ km s$^{-1}$. The synthesized beam located at bottom right is $10.8\times6.11$ arcsec$^2$, with a P.A. of 10.79$^{\circ}$. Black contours:  MAMBO-II 1.2 mm dust continuum emission at an angular resolution of 15~arcsec. Contours are 59, 79, 89, 99 percent of the peak, 57 mJy beam$^{-1}$.}
  \label{fig:Mom0}
\end{figure}
In Figure~\ref{fig:Mom0} we show the moment-zero map of the merged IRAM 30m and PdBI data (see Section~\ref{combined_obs}), overlapped with IRAM 30m MAMBO-II map of the dust continuum emission at 1.2 mm. The peak is 1 Jy beam$^{-1}$ km s$^{-1}$ and the arc structure hinted at the column-density single-dish map (panel (b) of Fig.\ref{fig:VESPAhfs}) is now clearly seen. 
\subsection{Chemical survey (FTS)} \label{FTSdata}
\begin{figure*}
 \centering
 \includegraphics[scale=0.8,keepaspectratio=true]{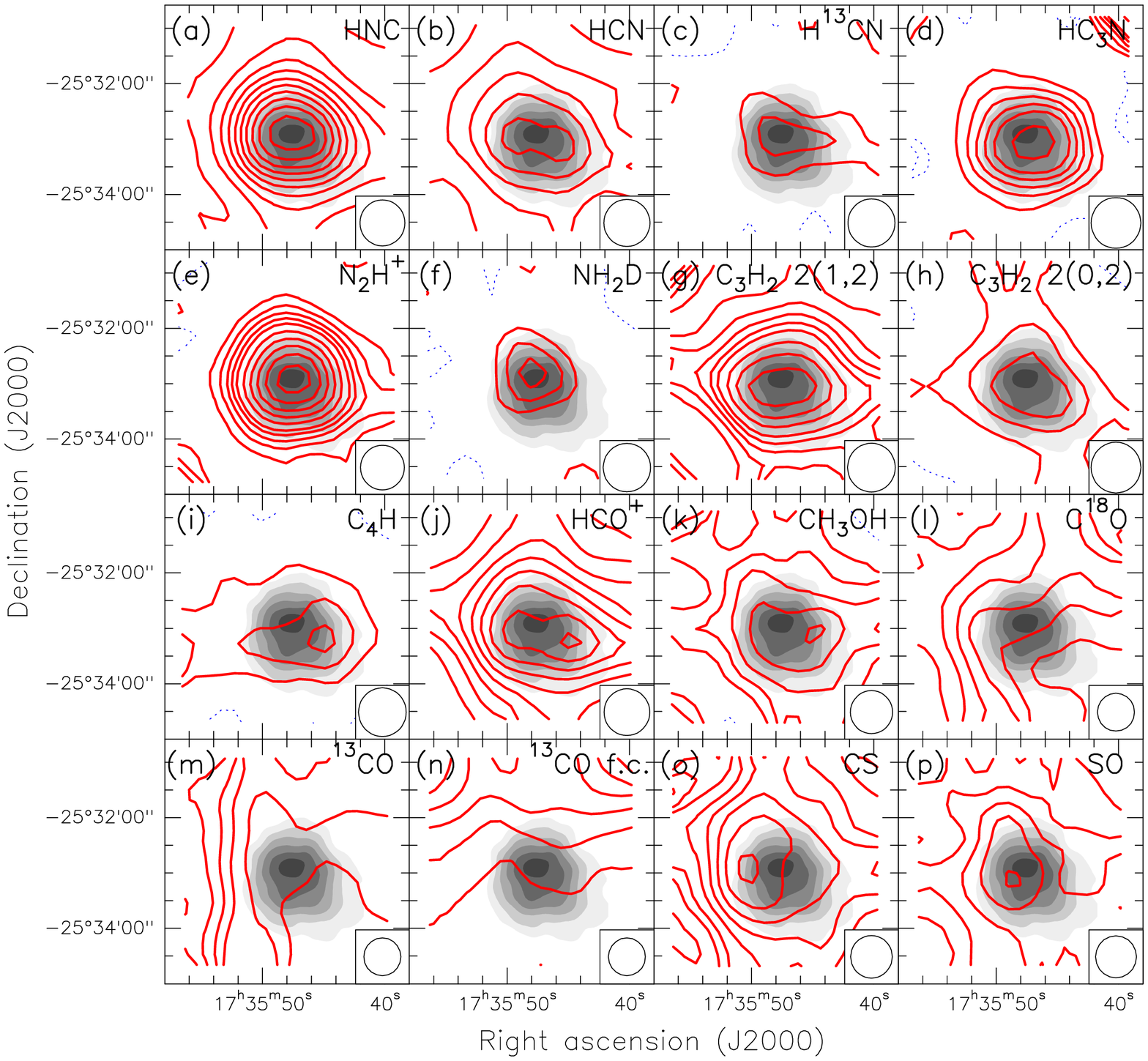}
  \caption{Red contours: averaged intensity maps of the observed molecular transition indicated in each panel. Contours in (a), (j) and (l) are 10, 20, 30, ... times the rms of the maps, 8, 6 and 9 mK, respectively. In panels (m) and (n) contours are 20, 23, 26, ... and $-3$, 3, 6, 9, ... times the rms, 53 mK, respectively. In panels (b) and (h), (k) and (p) countours are $-3$, 3, 6, 9, 12, 15, 20, ... times the rms, 21 mK and 16 mK, respectively. In the rest of the maps contours are $-3$, 3, 6, 9, 12, 15, 20, 25, 30, 35, ... times the rms, $10$ mK. Panel (n) shows a fainter component of the $^{13}$CO molecular line. The beam size of each molecular line is shown at the bottom right corner of each panel. Grey scale: IRAM 30m MAMBO-II map of the dust continuum emission at 1.2 mm at an angular resolution of 21~arcsec.}
  \label{fig:mapaver}
\end{figure*}
\begin{table}
\caption{ Chemical differentiation}
\begin{center}
\begin{tabular}{  c  c  c  c  c  }
\noalign{\smallskip}
\hline\noalign{\smallskip}
\hline
Molecule				&$\theta^\mathrm{a}$ ($''$)	&S/N 	&Peak shift$^\mathrm{b}$ ($''$) 	&$\sigma_\mathrm{s}^\mathrm{c}$ ($''$)  \\\hline
HNC			 		&49.8					&106		&7							&0.2	\\		
HCN			 		&51.0					&16		&21, 26 						&1.4	\\			
H$^{13}$CN			&52.1					&8		&10							&3.1	\\		
HC$_3$N				&55.0					&20 		&14	 						&1.3	\\	
N$_2$H$^+$			&48.4					&52		&10							&0.4	\\	
NH$_2$D				&52.5					&10          &7                             	        	        		&2.3	\\		
C$_3$H$_2$	2(1,2)	&52.8					&33		&13							&0.7	\\		
C$_3$H$_2$	2(0,2)	&55.0					&12		&33							&2.1	\\		
C$_4$H				&52.8					&9		&43							&2.6	\\		
HCO$^+$ 				& 50.6					&103		&43							&0.2	\\		
CH$_3$OH 			&46.6					&14		&45							&1.5	\\		
C$^{18}$O 			&41.0					&81		&119							&0.2	\\		
$^{13}$CO 			&41.0					&41		&58							&0.4	\\		
CS 					&46.2					&41		&33							&0.5	\\		
SO 					&45.4					&15		&24							&1.4	\\
\hline	
\end{tabular}
\begin{list}{}{}
\item[$^\mathrm{a}$]Beam size after a Gaussian smoothing of 3 pixels (pixel size $\sim13''$).
\item[$^\mathrm{b}$]Peak shift with respect to the 1.2~mm dust continuum peak.
\item[$^\mathrm{c}$]Position uncertainty given by $\sigma_\mathrm{s}=(\frac{4}{\pi})^{1/4}\times\frac{\theta}{\sqrt{8\mathrm{ln}2}}/(\mathrm{S/N})$ \citep{Reid88}. 
\end{list}
\label{tab:offsets}
\end{center}
\end{table}
%
In this section, we describe the maps of 15 molecular lines detected with the IRAM 30m FTS spectrometer towards FeSt\,1-457 (see Figure~\ref{fig:mapaver} and Table~\ref{tab:transitions}). To generate the maps we performed a gaussian smoothing of 3 pixels and we used the {\sc{map\_aver}} task of {\sc{greg}} software to average the intensity of only the channels with detected emission to obtain a better signal-to-noise ratio and to better characterize the structure of the faintest lines. As presented in previous works \citep{Frau10,Frau12b} we observed a rich chemistry with emission from early- (e.g., C$_3$H$_2$, HCN, CS) and late-time molecules (e.g., N$_2$H$^+$). For the nitrogen-bearing molecules (e.g., HNC and N$_2$H$^+$; panels (a) to (f)) the gas emission is very compact and the peak coincides well with the peak of the 1.2 mm dust continuum emission. A similar behavior is seen in C$_3$H$_2$ (panels (g) and (h)). However, for C$_4$H and molecules containing oxygen in our sample (e.g., HCO$^+$, CH$_3$OH; panels (i) to (k)) the emission is shifted $\sim44''$ towards the southwest with respect to the dust. An extreme case of this would be the C$^{18}$O and $^{13}$CO molecules, which present strong extended emission peaking to the southwest, at $\sim120''$ and $\sim60''$ of the dust continuum, respectively (panels (l) and (m)). For the molecules containing sulphur (CS and SO, panels (o) and (p)) the emission is located to the east with respect to the dust peak, $\sim35''$ and $\sim25''$, respectively. The peak offsets of the molecular lines with respect to the 1.2~mm dust continuum peak are comparable to the beam sizes of the gaussian smoothed integrated intensity maps ($\sim45''$; see Table~\ref{tab:offsets}) and therefore significant. Panel (n) shows the emission of a second (and fainter) velocity component of the $^{13}$CO emission at 3 km s$^{-1}$.

Regarding the intensity of the lines, apart from the $^{13}$CO and C$^{18}$O which do not trace the core well, the brightest lines are HNC, HCO$^+$ and N$_2$H$^+$. We found the weaker lines to be C$_4$H, NH$_2$D and H$^{13}$CN. This is in agreement with the previous observations by \citet{Aguti07}. However, they find the morphology of N$_2$H$^+$ more compact compared to our data. This difference could be due to the better S/N of our observations which allowed us to detect weaker extended emission. As the FTS backend has poor spectral resolution, the specific values of the line intensities cannot be compared with other works due to the spectral dilution. This does not affect the relative positions as the lines are narrow throughout the core. Note that the VESPA and FTS N$_2$H$^+$ maps present very similar structure (see Figs. ~\ref{fig:VESPAhfs}a and ~\ref{fig:mapaver}e).


\section{Analysis}
\subsection{The Bonnor-Ebert sphere model fit} \label{BEmodel}
A suitable model to fit the Pipe Nebula starless cores is the Bonnor-Ebert sphere
\citep{Bonnor56,Ebert55}. It describes a self-graviting, pressure-confined,
isothermal gas sphere in hydrostatic equilibrium. The density profile can
be derived by solving the Lane-Emden equation

\begin{equation}
\frac{1}{\xi^2}\frac{d}{d\xi}\left(\xi^2\frac{d\phi}{d\xi}\right)=e^{-\phi},
\label{eq_BE}
\end{equation}

\noindent where $\xi$ is the non-dimensional radius,

\begin{equation}
\xi=r\frac{\sqrt{4 \pi G \rho}}{c_s},
\label{eq_xi}
\end{equation}

\noindent and $\phi$ the logarithm of the density contrast,

\begin{equation}
\phi=\ln{\left(\frac{\rho_c}{\rho}\right)}.
\label{eq_denCont}
\end{equation}

\noindent $r$, $\rho$, and $\rho_c$ are the radius, volume density and central
volume density, respectively. The sound speed, $c_s$, is defined as 
$c_s=\sqrt{kT/\mu}$. $G$, $k$, $T$, and $\mu$ are the gravitational constant, the
Boltzmann constant, the temperature, and the mean molecular mass, assumed to be
2.33. Imposing  boundary conditions at the core center, forcing the density to
be $\rho_c$ and its derivative to be 0 (i.e., $\rho=\rho_c$ and $d\rho/dr=0$ at
$r=0$), Eq.~\ref{eq_BE} can be solved numerically. The last ingredient left is
the outer confining pressure, $P_{\rm out}$, exerted at the outer radius,
$R_{\rm out}$, from which one can derive the $\xi_{\rm max}$ parameter
($\xi_{\rm max}=(R_{\rm out}/c_s)\sqrt{4 \pi G \rho}$) that uniquely
characterizes the Bonnor-Ebert solution. The critical value is $\xi_{\rm max}=6.5$, which 
corresponds to $\left(\rho_c/\rho\right)_{\rm max}=14.1$. For values larger than
$\xi_{\rm max}$ the equilibrium is unstable to gravitational collapse.
\begin{figure}
 \centering
 \includegraphics[scale=0.3,keepaspectratio=true,angle=-90]{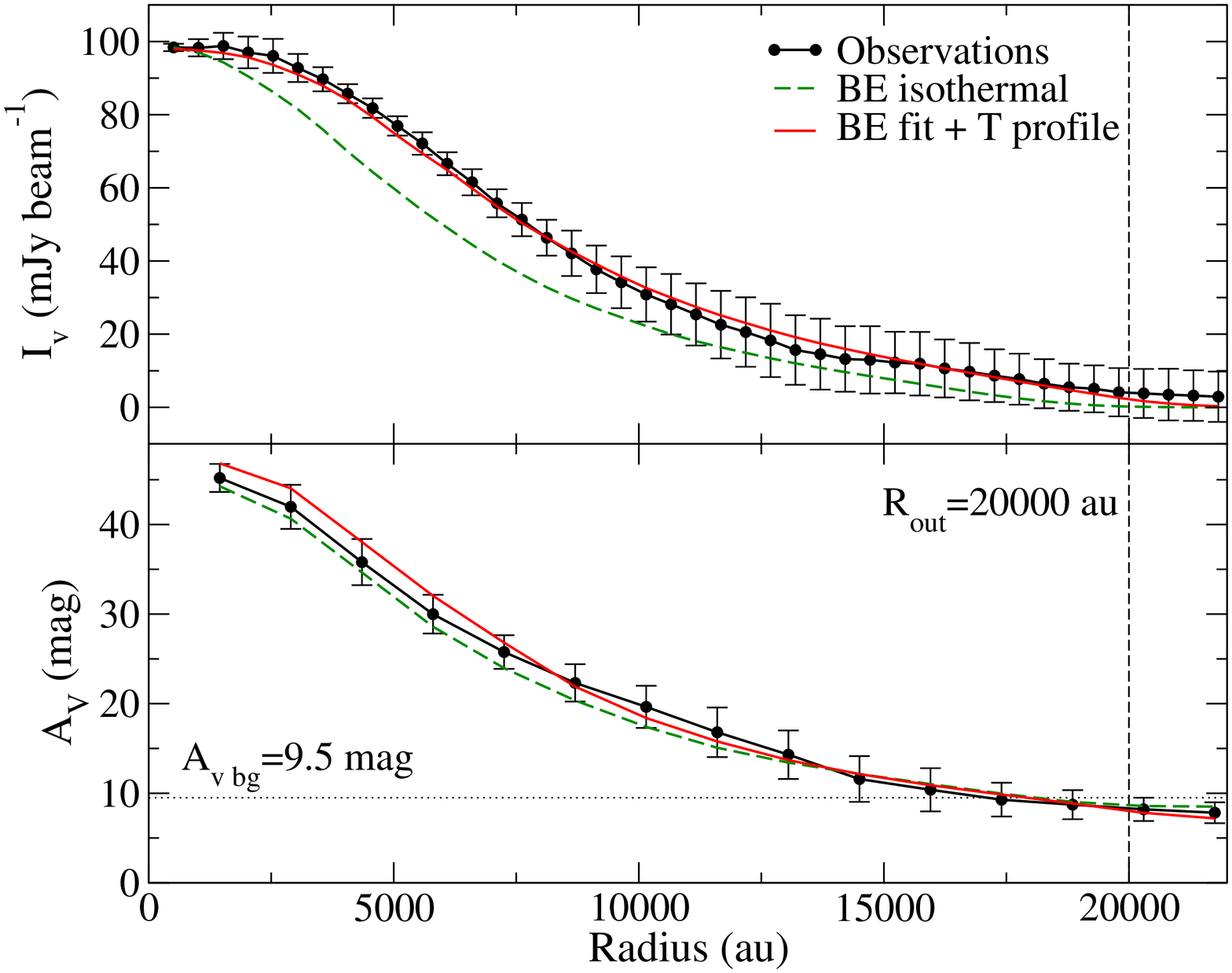}
  \caption{{\bf Upper panel:} Intensity profile from the 1.2 mm dust continuum emission map \citep{Frau10,Frau12b}. Black line and dots: observed values with vertical bars depicting the $\pm1$-$\sigma$ range. Green (long-dashed) line: isothermal Bonnor-Ebert fit. Red (solid) line: fitted Bonnor-Ebert profile adding the temperature profile shown in the upper panel of Figure~\ref{fig:BETn}. The best fitting central density is $\rho_c=2\times10^5$ cm$^{-3}$. The fitted core boundary radius, $R_{\rm out}=20000$ au is marked by a vertical dashed line.
{\bf Lower panel:} $A_\mathrm{V}$ profile from the dust extinction map \citep{Roman-Zuniga09,Roman-Zuniga10}. Black, green (long-dashed) and red (solid) lines as in upper panel. The fitted core boundary radius, $R_{\rm out}=20000$ au, and $A_\mathrm{V}$ value of the surrounding medium, $A_\mathrm{V}^{\rm bg}=9.5$ mag, are labeled and marked by a vertical dashed line and horizontal dotted line, respectively.}
  \label{fig:BE}
\end{figure}
\begin{figure}
 \centering
 \includegraphics[scale=0.3,keepaspectratio=true,angle=-90]{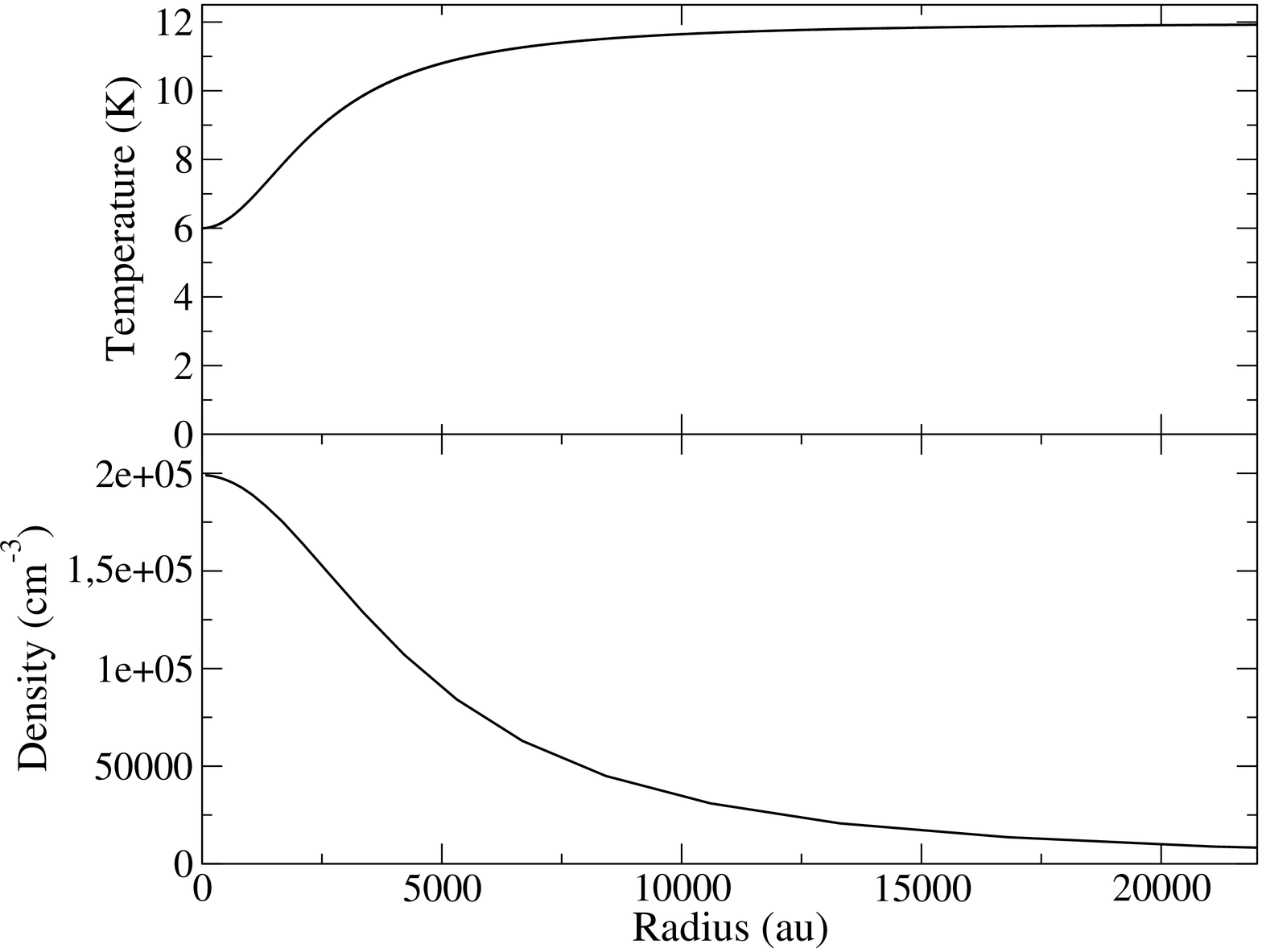}
  \caption{Temperature profile (upper pannel) and Bonnor-Ebert density profile (lower panel) used in this work (see Section~\ref{BEmodel}).}
  \label{fig:BETn}
\end{figure}
In order to fit the observational data of FeSt\,1-457, synthetic 3D cores were generated with density profiles obeying the numerical integration of a Bonnor-Ebert sphere (Eqs.~\ref{eq_BE}, \ref{eq_xi}, and \ref{eq_denCont}) with a given set of parameters. Then, the emission was integrated along the line-of-sight to generate 2D synthetic maps following a similar recipe as in \citet{Frau11}. We took into account the presence of the molecular cloud surrounding the core by adding a term in the extinction data, $A_{\rm v}^{\rm bg}$, which is constant along the extinction map. Since the IRAM 30m bolometer filters out the extended emission, we assumed that the molecular cloud does not contribute to the 1.2 continuum emission. Two maps were generated, a visual extinction map and a 1.2~mm dust continuum emission map, convolved to the respective observational beams \citep[$20''$ and $21.5''$ from][respectively]{Roman-Zuniga10,Frau10,Frau12b}. A dust opacity of  $\kappa_{\rm 250 GHz}=0.007$ cm$^2$~g$^{-1}$ was adopted for dust grains with thin ice mantles and for gas density $<10^6$ cm$^{-3}$ \citep{Ossenkopf94}. Finally, radial profiles of the observed and synthetic maps were obtained using the {\sc{ellint}} task from the {\sc{Miriad}} software \citep{Sault95}. This task averages the intensity in concentric annuli. The quality of the fit was assessed by computing the $\chi^2$ residual fit of the two radial profiles. Figure~\ref{fig:BE} shows the comparison between the observational (black lines and dots) and synthetic Bonnor-Ebert emission radial profiles (green (long-dashed) lines). 
\begin{table*}
\caption{Bonnor-Ebert fit parameters of FeSt\,1-457.}
\begin{center}
{\small
\begin{tabular}{lcccccccccc}
\noalign{\smallskip}
\noalign{\smallskip}
\hline
\hline
Model
&$\rho_c$
&$R_{\rm out}$
&$T^\mathrm{b}$
&$\xi_{\rm max}$
&$\rho_c/\rho_R$
&$P_{\rm out}$
&Mass
&$A_\mathrm{V}^{\rm bg}$
&Stable?
&Age$^\mathrm{c}$
\\
&($10^3$ cm$^{-3}$)
&($10^3$ au)
&($K$)
&
&
&($10^5$ $K$ cm$^{-3}$)
&(M$_\odot$)
&(mag)
&
&(yr)\\
\noalign{\smallskip}
\hline\noalign{\smallskip}
BE isothermal$^\mathrm{a}$		 &$200\pm20$	 &$20\pm1$	&$10.5\pm0.1$		&$10.5\pm0.7$	 &$47\pm5$	&$0.6\pm0.1$	&2.7	&9.5	&No	&$9.6\times10^5$ \\ 	
BE + T$_{\rm{profile}}^\mathrm{~d}$	 &$200\pm20$	 &$20\pm1$	&$12.0\pm0.1$		&$9.8\pm0.7$  	 &$40\pm5$	&$0.8\pm0.1$	&2.7	&9.5	&No	&$9.6\times10^5$ \\ 	
\hline
\end{tabular}
\begin{list}{}{}
\item[$^\mathrm{a}$]Parameters fitted for the isothermal BE model. The uncertainties of the fitted parameters $\rho_c$, $R_{\rm out}$ and $T$ were estimated by analyzing the increase of the $\chi^2$ residual fit by varying the parameters. The uncertainties of the derived physical parameters $\xi_{\rm max}$, $\rho_c/\rho_R$ and $P_{\rm out}$ were assessed through error propagation analysis.
\item[$^\mathrm{b}$]Temperature used to determine the sound speed value, $c_s$.
\item[$^\mathrm{c}$]Based on \citet{Aikawa05} results following the evolution of a marginally unstable Bonnor-Ebert sphere ($T=10$ $K$, $\alpha=1.1$, where $\alpha$ is the internal gravity to pressure ratio).
\item[$^\mathrm{d}$]Modified Bonnor-Ebert model using the temperature profile of the upper panel of Fig.~\ref{fig:BETn}. The uncertainties are the same as in the isothermal model because the new temperature profile does not modify the Bonnor-Ebert density profile, it only modifies the resulting 1.2~mm emission (see Section~\ref{BEmodel}).
\end{list}
}
\end{center}
\label{tab:BEfit}
\end{table*} 
The best fit was obtained for the values shown in Table~\ref{tab:BEfit}. The first row lists the fitted parameters of the Bonnor-Ebert profile
($\rho_c$, $R_{\rm out}$, and $T$), the physical parameters derived ($\xi_{\rm
max}$, $\rho_c/\rho_R$, $P_{\rm out}$, and mass), the fitted background visual
extinction arising from the surrounding ambient medium ($A_\mathrm{V}^{\rm bg}$), the
gravitational stability status, and an estimate of the age \citep{Aikawa05}. There is no degeneracy between the fitted parameters $\rho_c$, $R_{\rm out}$, and $T$ as the visual extinction map does not depend on temperature and  $\rho_c$ and $R_{\rm out}$ modify the height and width of the radial profile, respectively. The temperature only modifies the 1.2~mm dust continuum emission map. We found 
that the temperature derived from the Bonnor-Ebert fit ($10.5\pm0.1$~K) is in fair agreement with 
the temperature derived from ammonia observations \citep[$9.5\pm0.1$ K;][]{Rathborne08}. The value obtained for the derived parameter $\xi_{\rm max}$ was $10.5\pm0.7$. The error was estimated through error propagation analysis, being $\rho_c$ and $R_{\rm out}$ the main contributors to its uncertainty (see Table~\ref{tab:BEfit}). This value is significantly higher than the critical value (6.5), thus the core is gravitationally unstable, in accordance with \citet{Kandori05} results.

FeSt\,1-457 is observed with an excellent signal-to-noise to test the Bonnor-Ebert scenario. By using the isothermal Bonnor-Ebert sphere we found a good fit for the $A_\mathrm{V}$ map but non-negligible differences in the dust emission profile (see green (long-dashed) line in Figure~\ref{fig:BE}). This can be interpreted as a departure from the isothermal approximation because the extinction map ($A_\mathrm{V}$), contrary to the dust continuum emission, is only sensitive to the column density and does not depend on temperature. Previous studies in starless cores \citep[e.g.,][]{Crapsi07} show that those with $n_{H_2}>10^5$ cm$^{-3}$ may have an inner temperature gradient. These cores are colder at the center, reaching temperatures as low as 6 K, than at the outer shells, where they may be heated by the interstellar radiation field \citep[e.g.,][]{Marsh14}. We tested different temperature profiles to check whether the fits improved. We found that a temperature profile following: 
\begin{equation}
T(r)= T_{\mathrm{out}}-\frac{(T_{\mathrm{out}}-T_{\mathrm{in}})}{(1+(r / r_{\mathrm{in}})^2)}
\end{equation}
($r_{\mathrm{in}}=2500$ au) with $T_{\mathrm{in}}=6$ K at the center of the core and increasing until $T_{\mathrm{out}}=12$ K at a radius of 9000 au fitted much better the data. The second row of Table~\ref{tab:BEfit} shows the best fit parameters of the Bonnor-Ebert+temperature gradient model.
Figure~\ref{fig:BETn} shows the  Bonnor-Ebert density and temperature profiles that appear to fit best the data.  In addition, the red line in Figure~\ref{fig:BE} shows the predicted 1.2 mm dust intensity and extinction radial profile from this Bonnor-Ebert+temperature gradient model. \citet{Forbrich15} also find a clear decrease of the effective dust temperature towards the center of FeSt\,1-457 from Herschel observations. However, they find the lowest value at the center to be 13.5 K, higher than kinetic gas temperature derived from ammonia \citep[9.5 K;][]{Rathborne08} and higher than the temperature at the center of the core adopted by us. They argue this difference could be related to the effect of averaging the emission from Herschel along the line of sight which would take into account dust from the cloud at higher temperatures. Note that the temperature profile we found fits well the FeSt\,1-457 core is similar to the one found for the L1544 core \citep{Crapsi07}, a core with significantly higher density, but also to core 16 from \citet{Marsh14}, which has a density similar to that of FeSt1-457.

Even though we did not modify the Bonnor-Ebert density profile (lower panel of Figure~\ref{fig:BETn}), which assumes a constant temperature, we finally considered the temperature profile shown in the upper panel of Figure~\ref{fig:BETn} to generate the synthetic maps. This apparent inconsistency is not important in our case. \citet{Sipila11} show that the isothermal and non-isothermal Bonnor-Ebert density profiles are fairly similar. In addition, as the gas pressure is proportional to the temperature, reducing the temperature by half, reduces the pressure by half as well, which makes the core even more unstable against gravity. Therefore, this model is good enough to say that FeSt\,1-457 has a density profile of a very young core, it is gravitationally unstable and seems to be colder at the center than in the outer parts.
\subsection{ARTIST} \label{artist}
Once the physical model of temperature and density of the core was determined (see previous section), we estimated the abundances for the molecular lines detected with the IRAM 30m observations. 

We used the Adaptable Radiative Transfer Innovations for Submillimeter Telescopes \citep[ARTIST;][]{Brinch10,Padovani11,Padovani12,Jorgensen14} software to generate a synthetic spectral cube of the molecular lines detected. We assume that the source is spherically symmetric with the density and temperature radial profiles derived from the Bonnor-Ebert model derived in the previous section.  

For the simulations we assumed the dust temperature equal to the gas temperature. We assumed no velocity field in the core. The velocity gradient in the core is very small ($\la 0.05$~km~s$^{-1}$~beam$^{-1}$ at the center of the core), so this does not affect the results. Initially, the 
only free parameter used in ARTIST was the abundance. First, we assumed a constant value for the abundance. If the model did not fit well the observed intensity profile, we used a step function with two or three different abundances to obtain more accurate fits. We systematically tried different values for the radius at which there was a change in abundance, until obtaining the smallest $\chi^2$ parameter. One or two step functions (at a radius of $55$ arcsec ($\sim8000$ au) and $100$ arcsec (14500 au)) for the abundance was necessary to obtain good fits for 10 out of the 12 molecules, in which a constant abundance value did not fit well the observed data. We made use of the Leiden Atomic and Molecular Database (LAMDA)\footnote{http://home.strw.leidenuniv.nl/$\sim$moldata/} to load the collisional coefficients of the observed molecules \citep{Green74,Green78,Flower99,Chandra00,Daniel05,Lique06a,Lique06b,Rabli10,Yang10}. 
The spectral cube generated with ARTIST was convolved with a Gaussian to mimic the angular resolution of each molecular line observed. The output from ARTIST is given in units of brightness temperature.

In order to compare both the observed and the synthetic molecular line maps, we generated line intensity profiles for both data sets in the same way using the {\sc{Miriad}} software.   We first computed integrated intensity maps (moment-zero) using the {\sc{moment}} task   over a velocity range that included all the emission.  Then,  we used the {\sc{ellint}} task for the integrated intensity maps to compute the line radial profiles with annuli of 10~arcsec width, which is a value between one third and one fifth of the angular resolution of the observed molecular lines.

Figures~\ref{ellint4} and~\ref{fig:FTSabundance} show the annuli averaged integrated intensity (moment-zero) vs. the annuli radius for the observed (open circles) and ARTIST produced data. 
\subsubsection{N$_2$H$^+$ abundance fit} \label{Xn2hp} 
\begin{figure}
\begin{center}
\begin{tabular}[b]{c}
     \epsfig{file=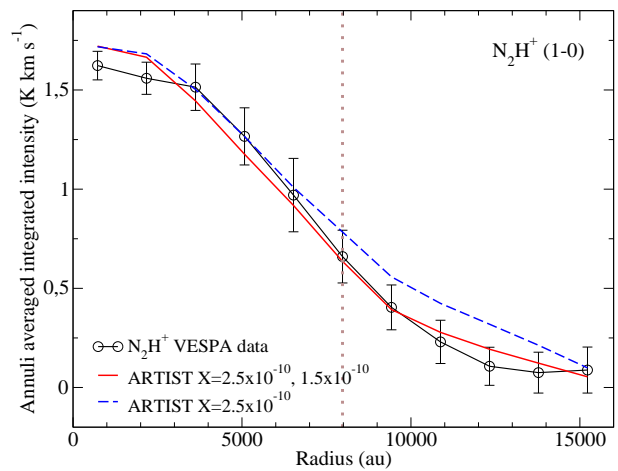, width=6.5cm,angle=-90} \\
\end{tabular}
\caption{N$_2$H$^+$ ARTIST abundance estimate. Annuli averaged integrated intensity vs. annuli radius.
Black line: IRAM 30m VESPA data. Blue dashed line: ARTIST radial integrated intensity profile for a constant abundance of $2.5\times10^{-10}$. Red line: ARTIST radial integrated intensity profile with a step function abundance of $2.5\times10^{-10}$ and $1.5\times10^{-10}$ for smaller and larger radius than $\sim8000$ au (55 arcsec; indicated by the brown vertical dotted line), respectively. 
}
\label{ellint4}
\end{center}
\end{figure}
To obtain the moment-zero map of N$_2$H$^+$ (see Fig.~\ref{fig:VESPAhfs}a), we only included the three central hyperfine lines to avoid the anomalous hyperfine lines (see Section~\ref{N2Hp30m}) and we obtained a good fit of the data. The high spectral resolution ARTIST fit is shown in the lower panels of Fig.~\ref{fig:hfsfits} \citep[note that ARTIST handles line overlapping:][]{Brinch10}. We first used a constant abundance (with respect to H$_2$) to fit the radial profile. A value of $2.5\times10^{-10}$ fitted well the data except for the outer radius where the model gave slightly higher values (blue dashed line of Fig.~\ref{ellint4}). Thus, we used a step function varying the radius at which there was a change in abundance systematically in steps of 10 arcsec, and a better fit was obtained with values of $2.5\times10^{-10}$ and $1.5\times10^{-10}$ for smaller and larger radius than 55 arcsec ($\sim8000$ au), respectively (red line of Fig.~\ref{ellint4}). The error of the radius where the abundance changes is $\pm15$ arcsec, estimated from a $\chi^2$ analysis \citep[e.g.,][]{Gibb00}. 
We also estimated the abundance uncertainty for the N$_2$H$^+$ line to be less than $\sim15\%$. This uncertainty was estimated from a $\chi^2$ analysis using different values of the abundance at the same range of radii of the core. 


\subsubsection{FTS molecular lines abundance fits} \label{FTSabundance}
\begin{figure*}
 \centering
  \includegraphics[scale=0.65,keepaspectratio=true,angle=-90]{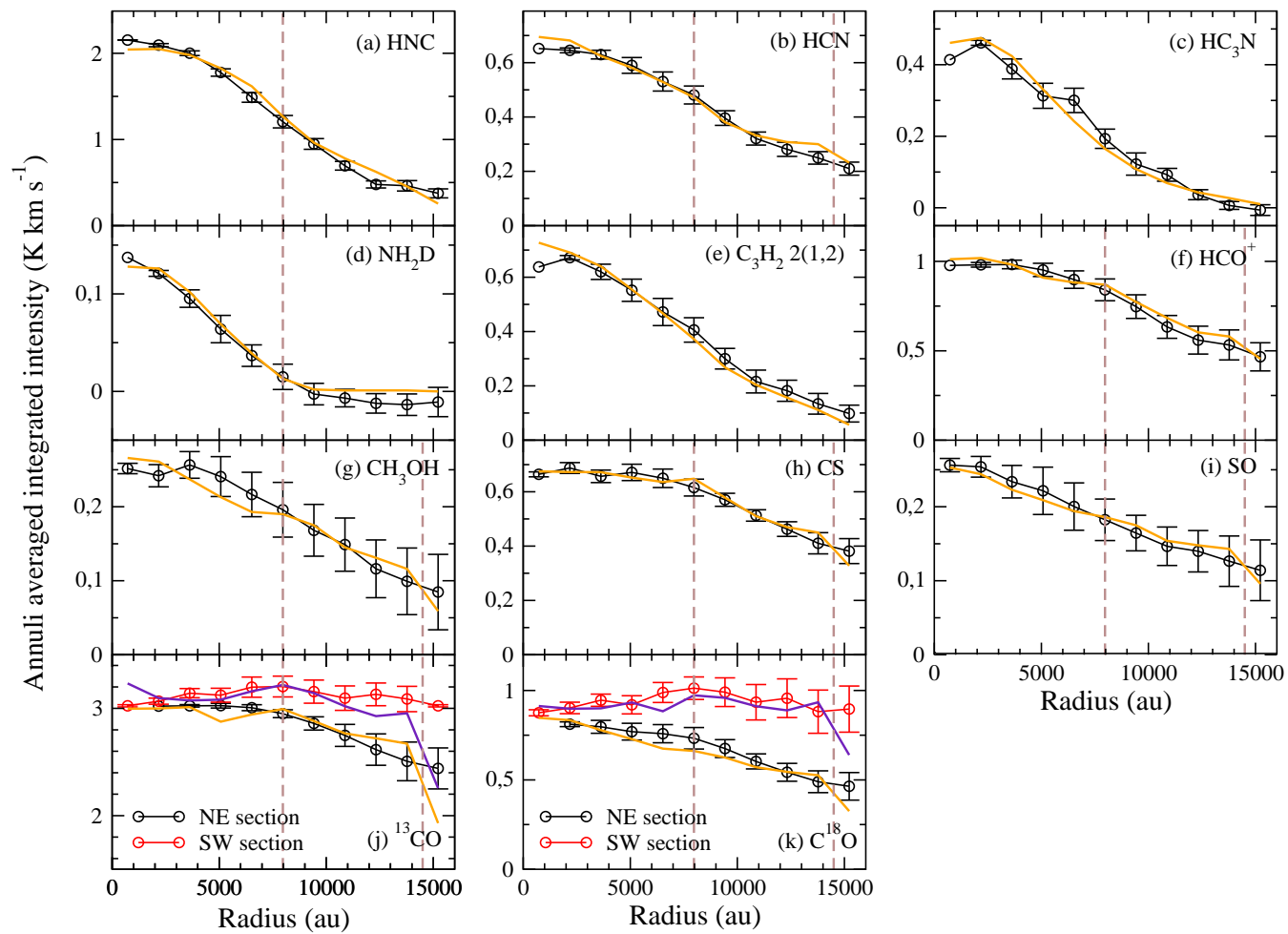}
  \caption{FTS molecular lines abundance fits. Annuli averaged integrated intensity vs. annuli radius. Black and red lines: FTS data profiles. Orange and violet lines: ARTIST data profiles with estimated abundances. Brown vertical dashed lines indicate the radius where there is a change in abundance (see Table~\ref{tab:Xvalues}).}
   \label{fig:FTSabundance}  
\end{figure*}
\begin{table}
\caption{Estimated molecular abundances using ARTIST.}
\begin{center}
\begin{tabular}{  c  c  c  c  }
\noalign{\smallskip}
\hline\noalign{\smallskip}
\hline
{Molecule}	&\thead{$0''-55''$\\$0-8000$\\(au)$^\mathrm{a}$}&\thead{$55''-100''$\\$8000-14500$\\(au)$^\mathrm{a}$}	&\thead{$>100''$\\$>14500$\\(au)$^\mathrm{a}$}\\\hline
N$_2$H$^+$			&$2.5\times10^{-10}$			&$1.5\times10^{-10}$				&$1.5\times10^{-10}$		\\
HNC			 		&$1.5\times10^{-9}$				&$6.0\times10^{-10}$				&$6.0\times10^{-10}$		\\
HCN			 		&$3.0\times10^{-10}$			&$4.0\times10^{-10}$				&$2.0\times10^{-9}$			\\
HC$_3$N				&$4.5\times10^{-10}$			&$4.5\times10^{-10}$ 				&$4.5\times10^{-10}$		\\
NH$_2$D				&$3.0\times10^{-10}$			&$<10^{-11}$                                    		&$<10^{-11}$ 				\\
C$_3$H$_2$			&$4.0\times10^{-10}$ 			&$4.0\times10^{-10}$				&$4.0\times10^{-10}$		\\
HCO$^+$ 				& $5.0\times10^{-11}$			&$1.5\times10^{-10}$				&$7.0\times10^{-10}$		\\
CS 					& $1.5\times10^{-10}$			&$8.0\times10^{-10}$				&$4.0\times10^{-9}$			\\
SO 					& $6.0\times10^{-11}$			&$2.0\times10^{-10}$				&$2.0\times10^{-9}$			\\
CH$_3$OH 			& $1.2\times10^{-10}$			&$2.5\times10^{-10}$				&$1.0\times10^{-9}$ 			\\
C$^{18}$O NE			& $1.5\times10^{-8}$				&$4.0\times10^{-8}$					&$2.0\times10^{-7}$			\\
C$^{18}$O SW			& $4.0\times10^{-9}$				&$6.0\times10^{-8}$					&$4.0\times10^{-7}$			\\
$^{13}$CO NE			& $8.0\times10^{-8}$				&$3.5\times10^{-7}$					&$2.0\times10^{-6}$			\\
$^{13}$CO SW			& $4.0\times10^{-8}$				&$4.0\times10^{-7}$					&$3.0\times10^{-6}$			\\
\hline	
\end{tabular}
\begin{list}{}{}
\item[$^\mathrm{a}$]Annuli radius.
\end{list}
\label{tab:Xvalues}
\end{center}
\end{table}
We followed the same procedure used for N$_2$H$^+$ to fit the other molecular lines detected with the FTS spectrometer, except for C$_4$H for which the collisional coefficients file was not available in the LAMDA catalogue. H$^{13}$CN and C$_3$H$_2$ 2(0,2)$-$1(1,1) were not fitted as they present weak emission and we had better signal to noise in the similar molecules/transitions HCN and C$_3$H$_2$ 2(1,2)$-$1(0,1). The observed intensity profiles are presented in Fig.~\ref{fig:FTSabundance} (open circles). Our ARTIST modeling indicated that some molecular lines are  well fitted with just one abundance value. This is the case of HC$_3$N and C$_3$H$_2$ 2(1,2) (see panels (c) and (e) of Fig.~\ref{fig:FTSabundance}). HNC and NH$_2$D are well fitted with a one step function, one value at the inner part of the core until 55 arcsec ($\sim 8000$ au) radius and a smaller value at larger radii (see panels (a) and (d) of Fig.~\ref{fig:FTSabundance}). However, our ARTIST model revealed that most of the lines needed 3 different abundance values changing at 55 and 100 arcsec ($\sim 8000$ and 14500 au) radius to obtain a good fit (radii where we assumed a change in abundance are marked in Fig.~\ref{fig:FTSabundance} by vertical dashed lines). 
 
We estimated the abundance uncertainty for the CH$_3$OH (one of the molecules with the lowest signal-to-noise ratio) and HCN (one of the molecules with the highest signal-to-noise ratio) following the same procedure as with the N$_2$H$^+$ giving similar results.

In all cases where we introduced a 2 step function profile in the abundance, values are higher in the outer part of the core and decrease towards the center. We consider a slightly different analysis for the abundances of C$^{18}$O and $^{13}$CO lines. As their emission is very asymmetric presenting their peak at the southwest part with respect to the dust continuum (see panels (l) and (m) in Fig.~\ref{fig:mapaver}), we divided the image in two sections, one including the strongest emission (southwest) and the other including the faintest emission to the northeastern part, which resembles more the emission pattern of the other species. Thus this part should trace more reliably the behavior of this molecule in the core. We performed the same procedure for the two sections separately. Both molecular lines show very flat profiles fitted with a 2 step function for the abundance as the emission is very extended (see panels (j) and (k) of Fig.~\ref{fig:FTSabundance}). Table~\ref{tab:Xvalues} summarizes the abundances for each molecule. Note that by using the main beam brightness temperature scale the intensity of the lines with extended emission, namely C$^{18}$O and specially $^{13}$CO, may be overestimated. This may result in some overestimation of their abundances but not of the observed relative drop within the core.

\subsection{Estimation of the magnetic field strength} \label{Chandrasekhar}
We made use of the polarization data presented in \citet{Alves14} to estimate the magnetic field strength on the plane of the sky. The observations were performed at the APEX 12m telescope at 345 GHz obtaining a FWHM of $\sim20''$. The magnetic field lines are relatively uniform with a weighted mean position angle of $130^{\circ}$ (from north to east direction).

Using the Chandrasekhar-Fermi technique we are able to estimate the magnetic field strength. We used the \citet{Ostriker01} approximation for estimates of the plane-of-sky field strength under strong-field cases ($\delta\phi\le25^{\circ}$). The projected field strength ($B_\mathrm{pos}$) is given by:
\begin{equation}
B_\mathrm{pos}=8.5\frac{\sqrt{n_{\mathrm{H_2}}/(10^6\mathrm{~cm}^{-3})}\Delta v/(\mathrm{km~s}^{-1})^{-1}}{\delta\phi \mathrm{~deg}^{-1}} \mathrm{~mG}, 
\end{equation}
where the number density of molecular hydrogen $n_{\mathrm{H_2}}$ is taken as $2.33m_\mathrm{H}n_\mathrm{H}$, $\Delta v$ is the observed line width and $\delta\phi$ is the dispersion of the polarization angles.

The observed dispersion $\delta\phi_{\mathrm{obs}}$, is produced by contributions from the measurement uncertainty of the polarization angle $\sigma_{\mathrm{PA}}$ and the intrinsic dispersion $\delta\phi_{\mathrm{int}}$. 
Taking into account every pair of polarization angles (PA) and their uncertainties $\sigma_{\mathrm{PA}}$ according to the equation 
\begin{equation}
\delta\phi_{\mathrm{int}}^2=\frac{\sum\limits_{i,j}\left(\mathrm{PA}_i-\mathrm{PA}_j\right)^2-\left(\sigma_{\mathrm{PA}_i}^2+\sigma_{\mathrm{PA}_j}^2\right)}{N},
\end{equation}
where $N$ is the number of data pairs, we obtained $\delta\phi_{\mathrm{int}}=9^{\circ}.22$ 

\citet{Frau10} obtain $n_{\mathrm{H_2}}=4\times10^5$ cm$^{-3}$, however, we adopted $n_{\mathrm{H_2}}=6\times10^4$ cm$^{-3}$, as \citet{Alves14} find depolarization at higher densities. From our N$_2$H$^+$ (1--0) observations, $\Delta v\sim0.2$ km~s$^{-1}$. Thus, we obtained $B_\mathrm{pos}\sim45.2~\mu$G and assumed to be uncertain by a factor of two.

\section{Discussion}
\subsection{Molecular depletion} \label{depletion} 
After our analysis with ARTIST (see Section~\ref{artist}), we found that molecular abundances are well fitted with three values, presenting a decrease of abundance towards the inner region of the core in most of the molecules detected (7 out of 12). The simplest explanation for the decrease of molecular abundance is that the molecules are depleted onto the dust grains once the density of the core increases and the temperature decreases. Some of these molecules are strongly depleted in the core interior with a decrease of abundances by at least 1 or 2 orders of magnitude. The strongest depletion is found in $^{13}$CO and C$^{18}$O with a difference in abundance of 2 orders of magnitude, from $\sim10^{-6}$ to $\sim10^{-8}$ and $\sim10^{-7}$ to $\sim10^{-9}$, respectively. \citet{Aguti07} find CO is depleted only by a factor of $\sim5$ but by averaging the abundance along the line of sight. CS and SO also present strong depletion with a decrease of abundance of a factor of $\sim30$ \citep[it is known that sulphurated molecules tend to deplete quite fast in chemical models of dense cores, e.g.,][]{Ruffle99}. Similar results are found in \citet{Tafalla02,Tafalla04} in several starless cores. We also found a decrease in abundance of 1 order of magnitude in HCN and HCO$^+$. The HCN shows emission peaks at both sides of the continuum peak in an east-west strip. The decrease of emission around the continuum peak is probably an indication of heavily depletion at that position. Curiously, the shape of the HCN and HNC emissions are quite different close to the continuum peak. \citet{Loison14} found that the HCN/HNC ratio tends to decrease when carbon atoms or CO are depleted, due to isomerisation of HCN into HNC. In fact, we see an increase of abundance of a factor of $\sim2$ of HNC at the inner region of the core. Thus, the combination of depletion of HCN plus the conversion of HCN into HNC could be enough to explain the observed difference in emission.

Our best studied molecule, N$_2$H$^+$, is well fitted with two abundance values with a small increase towards the interior of the core (see Section~\ref{Xn2hp}). The abundance value $\sim10^{-10}$ is in agreement with \citet{Aguti07} results. In Section~\ref{N2Hp30m}, we showed that the N$_2$H$^+$ column density follows an arc-structure around the 1.2 mm continuum peak (see Fig.~\ref{fig:VESPAhfs}b) which is clearly seen using the combined emission of IRAM 30m and PdBI (see Figure~\ref{fig:Mom0}). This might not be seen in the ARTIST analysis because we averaged the emission in concentric annuli and the feature is only present at a very small scale of 5 arcsec or 725 au, suggesting that N$_2$H$^+$ might be frozen-out onto the dust grains at this scale, which corresponds to a density of $\sim2\times10^5$ cm$^{-3}$. \citet{Aguti07} also suggest posible depletion of N$_2$H$^+$ in FeSt\,1-457 at depths >40 mag of visual extinction, which corresponds to radii $\la3350$ au and densities $\ga1.3\times10^5$ cm$^{-3}$. Our results are consistent with previous observations towards starless cores, where N$_2$H$^+$ is depleted at densities $>5\times10^5$ cm$^{-3}$ \citep[e.g.,][]{Belloche04,Pagani07}. The arc-like structure is not likely due to opacity effects as in the calculation of the column density we already took opacity into account. Higher angular resolution and with even better sensitivity observations are needed to discard other explanations for the arc structure seen in N$_2$H$^+$ (such as a real physical structure). 

NH$_2$D appears to have a significant abundance only in the inner region of the core, at radius $\la55$ arcsec ($\la8000$ au). Previous observational studies have shown evidence of NH$_2$D abundance increase towards starless cores \citep[e.g.,][]{Busquet10,Masque13}. Deuterated ammonia is expected to be abundant in the coldest and densest parts of starless cores, where the depletion of CO enhances the deuterium fractionation through a combination of gas phase \citep[e.g.,][]{Roberts04,Hatchell13} and surface reactions \citep[e.g.,][]{Fedoseev15,Fontani15}.

Finally, HC$_3$N and C$_3$H$_2$ are well fitted with a constant abundance value throughout the core. \citet{Tafalla06} study the molecular abundances of two starless cores (L1498 and L1517B) and find depletion of 1 order of magnitude of HC$_3$N towards the center of L1498 and of only a factor of 2 in L1517B. However, they find that HC$_3$N seems to survive in the gas phase to higher densities than other species as CO and SO. They also find a similar behavior in L1498 for C$_3$H$_2$ and suggest that the chemistry of C$_3$H$_2$ and HC$_3$N could be related. In our case, we can only see some depletion effect in the molecular profiles of HC$_3$N and C$_3$H$_2$ in the innermost position closer to the center of the core ($\la20$ arcsec, see panels (c) and (e) of Fig.~\ref{fig:FTSabundance}).

In summary, for most of the molecules detected in our sample (HCN, HCO$^+$, CH$_3$OH, CS, SO, $^{13}$CO and C$^{18}$O), we found a clear decrease in their abundance towards the inner region of the core.

\subsection{Chemical differentiation in the core's outskirts}
As we have seen in Section~\ref{FTSdata}, FeSt\,1-457 presents a clear chemical differentiation for nitrogen, oxygen and sulphurated molecules. Their emission appears with different offsets with respect to the peak of the 1.2 mm dust continuum emission (see Section~\ref{FTSdata} and Table~\ref{tab:offsets}). 
 
We computed the CS/C$^{34}$S, SO/$^{34}$SO and $^{13}$CO/C$^{18}$O line intensity ratios from the integrated spectra of an area of 30 arcsec radius around the position where the main isotopologue line is brighter. Because these ratios are done for the same rotational transition, they depend only on the optical depth and abundance ratio of the isotopologues \citep[we adopted an abundance ratio of 5.6 for $^{13}$CO/C$^{18}$O and 24 for $^{32}$S/$^{34}$S:][]{Wilson92,Chin95}.  The optical depth for CS (2$-$1), SO 3(2)$-$2(1) and $^{13}$CO (1$-$0) transitions is 1.54, 2.11 and 2.14, respectively. We expect this value to decrease as we go away from the peak. Thus, the observed different morphologies are not likely due to opacity. In addition, the critical densities and energies of the upper level of N$_2$H$^+$ (1$-$0), CS (2$-$1) and HCO$^+$ (1$-$0) are similar,  $\sim10^5$ cm$^{-3}$ and $4-7$ K, respectively (from LAMDA database; see Table~\ref{tab:transitions}). Therefore, we can exclude different excitation conditions as well. This suggests that different types of molecules are tracing different chemical conditions in the core. 

The characteristic asymmetries in the emission of C$_4$H, HCO$^+$ and CH$_3$OH, peaking (south)west of the dust peak, could be evidence of the penetration of UV radiation into  FeSt\,1-457. 

Moreover,  H$^{13}$CN and HC$_3$N also show a slight displacement of their emission peak towards the southwest. 
This may be a similar situation to the emission enhancement of these molecules in PDRs (photo-dominated regions) and in molecular condensations ahead of Herbig-Haro objects \citep[e.g.,][]{HT97,Girart98,Girart02,Viti06}. In the second case, chemical models show that this enhancement is due to the UV radiation arising from the Herbig-Haro objects \citep{Viti99,Viti03}. 
\citet{Viti03} find that the UV radiation affects the chemistry enhancing HCO$^+$ and CH$_3$OH for $A_\mathrm{V}<4$ mag, meaning that the enhanced emission should trace the most outer layer of the core.   Additionally,  PDR models show that C$_4$H  reach higher abundances at somewhat less deep layers (lower $A_\mathrm{V}$ values) than HCN.  Indeed, C$_4$H has been found  to peak close to the edge of PDRs as its production is enhanced by the UV radiation field in a relatively narrow range of visual extinction values \citep[e.g.,][]{Pety05,Morata08,Rimmer12}.  Moreover,  in some PDR models \citep[e.g.,][]{Morata08,Rimmer12}, HC$_3$N also has an abundance peak at similar layers to HCO$^+$.


FeSt\,1-457 is, in projection, embedded in a diffuse molecular cloud with average $A_{\rm V}^{\rm bg} \sim9.5$ mag. Thus, the enhanced C$_4$H, HCO$^+$ and CH$_3$OH emission could only happen if the core is near the border of the molecular cloud, such that one side of the core is exposed to the interstellar UV radiation without much absorption. We do not expect the penetration of the UV photons to be too large, but it is probably strong enough to produce some differences in the distribution of the molecular emission. Infrared extinction maps show that FeSt\,1-457 is surrounded by gas with relatively low extinction, $A_\mathrm{V}=5-10$ mag \citep{Roman-Zuniga10}. If the medium around FeSt\,1-457 is not homogeneous, some UV radiation from a nearby OB star or the interstellar radiation field could be penetrating quite deep into the (western side of the) core. Indeed, \citet{Gritschneder12} proposed that the Pipe Nebula might be an H~\textsc{II} region shell swept by the nearby $\theta$ Ophiuchi (a B2 IV $\beta$ Cephei star). $\theta$ Ophiuchi lies in a direction approximately west-northwest of FeSt\,1-457 and might be the source of the UV radiation required to explain the morphology of the emission of C$_4$H, HCO$^+$ and CH$_3$OH. Future works must be done to confirm this possible scenario.
\begin{figure*}[t]
 \centering
 \includegraphics[scale=0.5,keepaspectratio=true]{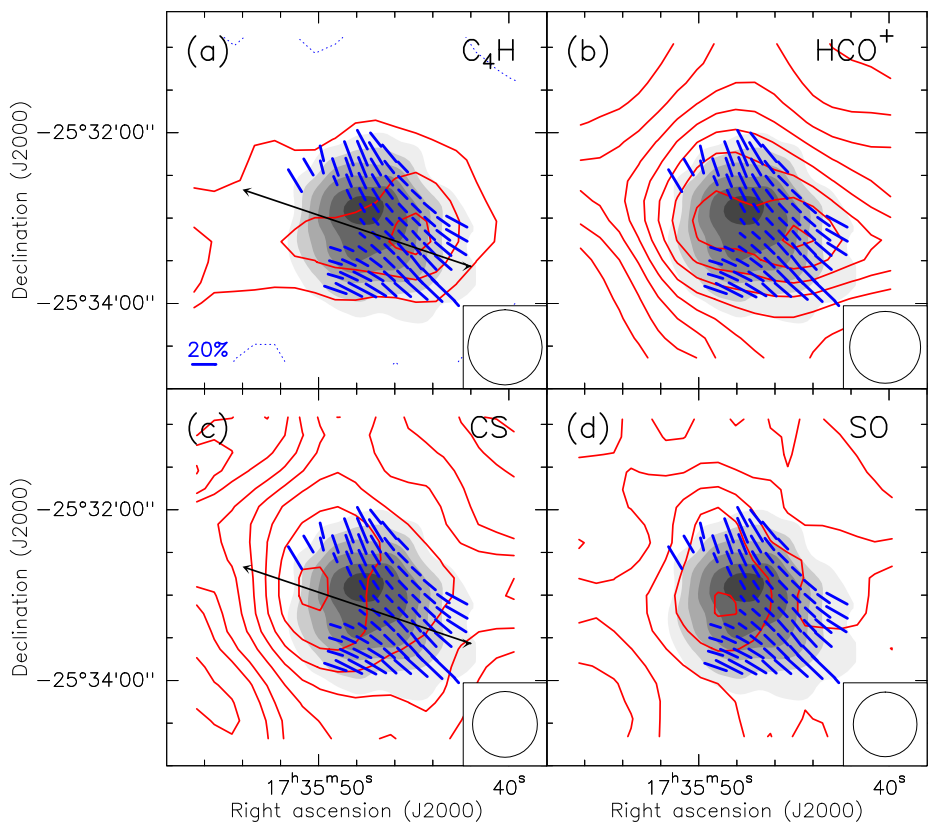}
  \caption{Polarization maps. Red contours: averaged intensity maps of (a) C$_4$H, (b) HCO$^+$, (c) CS and (d) SO. Contours in panels (a) and (c) are $-3$, 3, 6, 9, 12, 15, 20, 25, 30, ... times the rms of the maps, $10$ mK. Contours in panel (b) and (d) are 10, 20, 30, ... and $-3$, 3, 6, 9, 12, 15 times the rms of the maps, 6, and 16 mK. The beam size of each molecular line is shown at the bottom right corner of each panel. Grey scale: IRAM 30m MAMBO-II map of the dust continuum emission at 1.2 mm. Blue segments: Position angle of the linearly polarized data from \citet{Alves14}. The length of the segments is proportional to the polarization fraction. Black arrows in panels (a) and (c) indicate the 1-D slice shown in Fig.~\ref{fig:cgslice}.}
   \label{fig:dustpol}  
\end{figure*}

Regarding the asymmetries of CS and SO emission (see panels (o) and (p) in Fig.~\ref{fig:mapaver}, respectively), it has been studied that in PDRs \citep[see][]{Morata08} CS has a relatively shallow abundance closer to the edge of the PDR (around or inwards the C$_4$H peak) and a higher abundance in inner layers, while SO has high abundances in even deeper layers into the PDR. The combination of depletion and the influence of the UV field could explain the shape of the emission of these maps. If the eastern side of the core is more shielded from the radiation, these molecules will reach higher abundances and have their emission peaks there, while they will have lower abundances on the western side of the core (more markedly for SO), and they will be depleted towards the center of the core.
Finally, N$_2$H$^+$ and deuterated molecules, from both the observational and chemical point of view, are supposed to be abundant in the inner and most shielded parts of the cloud \citep[e.g.,][]{Roberts03,Crapsi05}, which is what our emission maps show (see Fig.~\ref{fig:mapaver} and~\ref{fig:FTSabundance}).
 
\subsection{A correlation between depletion, grain growth and polarization}
As we have seen in Section~\ref{artist}, and discussed in Section~\ref{depletion}, molecular abundances are well fitted with two or three values, presenting a decrease of abundance towards the inner region of the core in most of the molecules detected. This change in the chemical properties suggests FeSt\,1-457 experiences a variation in the physical properties of its inner region. 

It is interesting to note that \citet{Forbrich15} find evidence of grain growth at the inner region of the core from Herschel-derived dust opacity results and near-infrared extinction measurements. Depletion of the molecular gas onto the dust grains due to a decrease of temperature and increase of density in the center of the core could make easier for the grains to stick together and contribute to the grain growth seen by \citet{Forbrich15}. This is supported by the fact that we found a depletion radius very similar to the radius where Forbrich et al. find grain growth. It has been studied that water ice on the surface of dust particles enhance grain-grain adhesion favoring grain growth \citep[e.g.,][]{Gundlach15,Wang05}. Furthermore, \citet{Liu13} find that CO depletion positively correlates with a decrease of the dust emissivity index, which could be related to the size of the dust grains \citep[e.g.,][]{Lommen07,Ricci12,Pietu14}. This also agrees with our results of low abundance of CO at the center of the core (see Fig.~\ref{fig:mapaver}) and the grain growth seen by \citet{Forbrich15}.

In a different work, \citet{Alves14} present a study of the polarization properties in FeSt\,1-457 at different wavelengths and find that the emission is more polarized to the western side of the core \citep[see][figure 1]{Alves14}. It is globally accepted that non-spherical dust grains are aligned perpendicularly with respect to magnetic field lines producing linearly polarized thermal continuum emission \citep{Davis51,Hildebrand88}. One of the most important theories for the dust grain alignment is via radiative torques \citep{Dolginov76,Draine96,Draine97,Lazarian07}. Thus, the stronger polarized emission to the west also supports the idea of a UV radiation field penetrating into the core from the west which would be affecting not only the polarization properties but also the chemistry in the core (see previous section). 

\begin{figure}[t]
 \centering
 \includegraphics[scale=0.3,keepaspectratio=true,angle=-90]{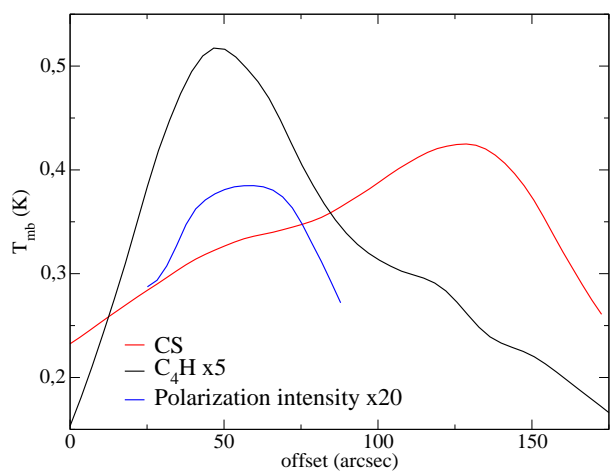}
  \caption{Intensity from CS (red), C$_4$H (black) and polarization intensity (blue) along a 1-D slice from the southwest part of the core towards the east passing through the C$_4$H and CS peaks (see panels (a) and (c) of Fig.~\ref{fig:dustpol}). Both C$_4$H and the polarization intensity increase at the same positions in the core along the slice. CS shows an anti-correlation with the polarization intensity.}
   \label{fig:cgslice}  
\end{figure}
In Fig.~\ref{fig:dustpol} we show the emission of 4 molecules peaking to the southwest (C$_4$H and HCO$^+$) and to the east (CS and SO) of the dust peak, overlaid with the position angles of the polarization data from \citet{Alves14}. The C$_4$H and HCO$^+$ emission seem to correlate with the stronger polarized emission towards the west. On the other hand, the emission peaks of CS and SO seem to present an anti-correlation with the polarized emission. To study further these possible correlations, we performed a 1-D slice across the core passing through the C$_4$H and CS peaks (black arrows in Fig.~\ref{fig:dustpol}). In Fig.~\ref{fig:cgslice} we show the polarization intensity (blue) and the emission of C$_4$H (black) and CS (red) along the slice. The polarization intensity presents a similar distribution and seem to correlate well with the C$_4$H emission and anti-correlate with the CS emission. To quantify these correlations we performed the statistical tests Kendall's $\tau$ and Spearman's $\rho$\footnote{http://www.statsoft.com/textbook/nonparametric-statistics/} and found a good correlation ($\tau=0.647$, $\rho=0.804$) between C$_4$H emission and polarization intensity and anti-correlation values ($\tau=-0.062$, $\rho=-0.058$) between CS and polarization intensity. More observations are necessary to further study these results.

On the other hand, \citet{Alves14} also find a steep decrease in the polarization efficiency at visual extinctions ($A_\mathrm{V}$) larger than 30 mag. \citet{Alves14} propose the lack of an internal source of radiation results in a loss of grain alignment with the magnetic field, which is consistent with the theory of dust grain alignment via radiative torques. Our work suggests that depolarization could be produced as a combination of lack of internal source and dust grain growth possibly affected by molecular depletion. Due to the `grain clustering' yielding to larger grains, the elongated shape of the grains could be distorted. The paramagnetic nuclei of the grains would be surrounded by a thick layer of ice (unpolarized) and the alignment of the grains could be harder to achieve resulting in a depolarized inner region as seen by \citet{Alves14}. 

Therefore, we conclude that depletion, grain growth and depolarization seem to be correlated in FeSt\,1-457.

\subsection{Magnetic properties of FeSt\,1-457}
Previous studies in star forming cores have shown that magnetic fields can play an important role in their evolution \citep[e.g.,][]{Girart06}.

Magnetic forces could prevent a cloud from collapsing if its mass-to-flux ratio is below the critical value,
\begin{equation}
M/\Phi_B<(M/\Phi_B)_{\mathrm{cr}},
\end{equation}
where M is its mass and $\Phi_B=\pi<B>R^2$ is the magnetic flux through it, derived from the mean magnetic field and the radius of the cloud cross-section perpendicular to the magnetic field. The critical mass-to-flux ratio is $(M/\Phi_B)_{\mathrm{cr}}=1/(2\pi G^{1/2})$ \citep{Nakano78}, where $G$ is the gravitational constant. We used the ratio in terms of the column density following eq.(7) of \citet{Pillai15}
\begin{equation}
\frac{(M/\Phi_B)}{(M/\Phi_B)_{\mathrm{cr}}}=0.76 \left(\frac{N_{H_2}}{10^{23}~\mathrm{cm}^{-2}}\right)\left(\frac{B_{\mathrm{tot}}}{1000~\mu G}\right)^{-1}.
\end{equation}
\citet{Frau10} obtain a core+cloud column density $N_{H_2}=47.6\times10^{21}$ cm$^{-2}$. From our Bonnor-Ebert model fit (see Section~\ref{BEmodel}) we found the background visual
extinction arising from the surrounding medium ($A_\mathrm{V}^{\rm bg}$) to be 9.5 mag which corresponds to a column density of $8.9\times10^{21}$ cm$^{-2}$. Thus, we adopted a column density for the core of $38.7\times10^{21}$ cm$^{-2}$. On the other hand, we assumed $B_{\mathrm{pos}}\sim B_{\mathrm{tot}}\sim45.2~\mu$G as \citet{Alves14} find the magnetic field direction mainly onto the plane-of-sky. We obtained a ratio of 6.5 (magnetically supercritical) which implies that gravity dominates over the magnetic support. Our Bonnor-Ebert model fit showed that the core is gravitationally unstable (see Section~\ref{BEmodel}). Finally, we can see that the magnetic field is not enough to avoid the collapse.

\section{Conclusions}
We have used the IRAM 30m telescope and the PdBI to study the chemical and physical properties of the starless core FeSt\,1-457 (Core 109) in the Pipe nebula. Our main conclusions are as follows:
\begin{enumerate}
\item We fit the hyperfine structure of the N$_2$H$^+$ (1$-$0) IRAM 30m data. This allows to measure with high precision the velocity field and line widths in the core. The N$_2$H$^+$ emission shows a clear southwest-northeast velocity gradient of 1.78 km s$^{-1}$ pc$^{-1}$, and larger line widths (0.3 km s$^{-1}$) at the southwest part of the map. The typical line widths are $0.17-0.23$ km s$^{-1}$. The column density map presents an arc-structure around the  the 1.2 mm dust continuum peak.
\item Combining both IRAM 30m and PdBI N$_2$H$^+$ (1$-$0) data we can resolve the arc-like structure hinted at the single dish map.
\item The core presents a rich chemistry with emission from early (C$_3$H$_2$, HCN, CS) and late-time molecules (e.g., N$_2$H$^+$), with a clear chemical spatial differentiation for nitrogen (centrally peaked), oxygen (peaking to the southwest) and sulphurated molecules (peaking to the east).
\item The chemical difference in the core could be due to an external UV radiation field penetrating into the core from the (south)west which could be also affecting the polarization properties. This implies that the core should be close to the edge of the molecular cloud.
\item FeSt\,1-457 is well fitted with a Bonnor-Ebert sphere model and introducing a temperature gradient decreasing towards the center from 12 to 6 K. We found the core is gravitationally unstable and that the magnetic field is not enough to stop the collapse.
\item We have analyzed the abundances of the molecular lines using the ARTIST software. For most of the molecules detected (HCN, HCO$^+$, CH$_3$OH, CS, SO, $^{13}$CO and C$^{18}$O), abundances are best fitted with three values, presenting a clear decrease of abundance of at least 1 or 2 orders of magnitude towards the inner region of the core. This is consistent with chemical models of starless cores, that show that there is a significant depletion of molecules onto the dust grains. On the other hand, N$_2$H$^+$, HC$_3$N and C$_3$H$_2$ are well fitted with a constant abundance throughout the core and the abundance of HNC and NH$_2$D increases in the inner region of the core (at radius $\la55$ arcsec). 
\end{enumerate}
Finally, we have seen that depletion of molecules onto the dust grains, grain growth and depolarization take place at the inner region of the core. This strongly suggests that these properties could be correlated in FeSt\,1-457. 

\begin{acknowledgements}
We would like to thank the referee for her/his useful comments. We also thank Phil Myers and Jan Forbrich for thoughtful discussions. CJ, JMG and RE acknowledge support from MICINN (Spain) AYA2014-57369-C3 grant. CJ and RE also acknowledge
MDM-2014-0369 of ICCUB (Unidad de Excelencia `Mar\'{\i}a de Maeztu'). JMG also acknowledges the support from the MECD (Spain) PRX15/00435 travel grant and from the SI CGPS award ``Magnetic Fields and Massive Star Formation''. AP acknowledges financial support from UNAM-DGAPA-PAPIIT IA102815 grant, Mexico. OM is supported by the MOST (Ministry of Science and Technology,
  Taiwan) ALMA-T grant MOST 103-2119-M-001-010-MY to the Institute of
  Astronomy \& Astrophysics, Academia Sinica.
\end{acknowledgements}


%
   \bibliographystyle{aa} 
   \bibliography{bibliography} 
%

\end{document}